\documentclass[times]{elsarticle}
\usepackage[utf8]{inputenc}

\usepackage[margin = 1in]{geometry}
\usepackage{booktabs}
\usepackage{amsmath}
\usepackage{hyperref}
\usepackage{amssymb}
\hypersetup{colorlinks=true, linkcolor=blue, citecolor=blue, urlcolor=blue}
\usepackage[]{natbib}
\usepackage{soul}
\usepackage{multirow}
\usepackage{diagbox}
\usepackage{comment}
\usepackage{float}

\newcommand{\bs}[1]{\boldsymbol{#1}}

\newcommand{\uvec}[1]{\underline{#1}}
\newcommand{\mat}[1]{\uvec{\uvec{#1}}}
\date{\today}

\begin{document}

\begin{frontmatter}
    \title{Computation of effective elastic moduli of rocks using hierarchical homogenization}

    \author[label1]{ Rasool Ahmad \corref{cor1}}
    \ead{rasool@stanford.edu}
    \author[label2]{ Mingliang Liu }
    \ead{mliu9@stanford.edu}
    \author[label3]{Michael Ortiz}
    \ead{ortiz@aero.caltech.edu}
    \author[label2]{Tapan Mukerji}
    \ead{mukerji@stanford.edu}
    \author[label1]{Wei Cai}
    \ead{ caiwei@stanford.edu }
    \address[label1]{Micro and Nano Mechanics Group, Department of Mechanical Engineering,
        Stanford University, CA 94305, USA}
    \address[label2]{Department of Energy Resources Engineering, Stanford University, CA 94305, USA}
    \address[label3]{Division of Engineering and Applied Science, California Institute of Technology, CA 91125 Pasadena, USA}
    \cortext[cor1]{corresponding author}

    \begin{abstract}
        This work focuses on computing the homogenized elastic properties of rocks from 3D micro-computed-tomography (micro-CT) scanned images. The accurate computation of homogenized properties of rocks, archetypal random media, requires both resolution of intricate underlying microstructure and large field of view, resulting in huge micro-CT images. Homogenization entails solving the local elasticity problem computationally which can be prohibitively expensive for a huge image. To mitigate this problem, we use a renormalization method inspired scheme, the hierarchical homogenization method, where a large image is partitioned into smaller subimages. The individual subimages are separately homogenized using periodic boundary conditions, and then assembled into a much smaller intermediate image. The intermediate image is again homogenized, subject to the periodic boundary condition, to find the final homogenized elastic constant of the original image. An FFT-based elasticity solver is used to solve the associated periodic elasticity problem. The error in the homogenized elastic constant is empirically shown to follow a power law scaling with exponent $-1$ with respect to the subimage size across all five microstructures of rocks. We further show that the inclusion of surrounding materials during the homogenization of the small subimages reduces error in the final homogenized elastic moduli while still respecting the power law with the exponent of $-1$. This power law scaling is then exploited to determine a better approximation of the large heterogeneous microstructures based on Richardson extrapolation.
    \end{abstract}

\end{frontmatter}

\section{Introduction}
\label{sec_intro}

Reliable prediction of properties of rocks from the pore-scale micro-computed-tomography (micro-CT) images is the defining aim of digital rock physics (DRP)~\citep{Fredrich1993, Arns2001,Dvorkin2011, Andra2013, Andra2013a, Saxena2017, Saxena2019, Sun2021}. The workflow of DRP starts with the acquisition of pore-scale rock 3D images from micro-CT scans. The scanned images are next subjected to image processing tools to segment them into underlying constituent materials and pores. The relevant physical simulations are then performed to compute the target physical properties of interest such as elastic constants, electrical conductivity, permeability, etc. This work is concerned with computing the effective  elastic properties (bulk and shear moduli) from a given segmented micro-CT image of a rock sample.

A rock is an archetypal example of random heterogeneous solids. The existing theoretical tools to deal with the homogenization of heterogeneous materials provide rigorous bounds on the homogenized elastic constants. The well known examples of such bounds include the Voigt-Reuss~\citep{voigt1910lehrbuch, Reuss1929} and Hashin-Shtrikman bounds~\citep{Hashin1963}. However, these bounds are generally quite loose and obtained by assuming certain idealized distribution of the phases constituting the heterogeneous materials. Thus, these rigorous theoretical bounds often fail to provide accurate estimates of microstructure-specific elastic constants of a heterogeneous material. For a comprehensive review of the homogenization of the elastic constants, readers are referred to Refs.~\citep{nemat2013micromechanics, Zaoui2002, Charalambakis2010}.

Computational homogenization techniques~\citep{yvonnet2019computational,Khdir2013, Takano2000, Vel2016} are often used to obtain the microstructure-specific effective elastic constants of materials. The partial differential equation (PDE) of elasticity (originated from equilibrium, compatibility conditions and material's constitutive relations) is solved numerically (e.g. using finite element method or fast fourier transform solvers) to find the stress and strain fields under a prescribed load.  The relationship between the averaged stress and strain values provides information on the effective elastic constants. A major issue in carrying out the computational homogenization arises from the sheer size of the rock sample images that is required for determining the well-converged effective elastic properties. Rock microstructures contain intricate multiscale features, i.e. the spatial distribution of minerals and pores that span over a range of length scales. The accuracy of the computationally homogenized properties depends on both the resolution and the field of view of the digital image~\citep{Saxena2019}.  A high resolution is needed to obtain a faithful representation of the fine microstructure, while a large field of view is necessary to ensure the sample is statistically representative. Unfortunately, refining the resolution of the scanned images while maintaining a large field of view would result in a very large the number of voxels in the digital image of the rock sample. The computational cost to perform physical simulations on these huge images quickly becomes prohibitive in terms of time and memory requirements and undermines the fundamental premises of DRP.

To alleviate the constraints of the available computational resources, here we apply a hierarchical homogenization procedure to compute the homogenized elastic constants that obviate the need of solving the elasticity equations on a given huge image by brute force. The basic idea is inspired by the renormalization method~\citep{Kadanoff1966,Wilson1971,Wilson1975,Wilson1983,Fisher1998,Efrati2014}, where we progressively scale down the size of the problem domain by successive applications of a coarse-graining rule at each scale. In the hierarchical homogenization scheme, a large rock image  is partitioned into smaller subimages of size $n\times n\times n$. Each smaller subimage is then homogenized computationally and replaced by an equivalent voxel with effective elastic constants. These voxels are subsequently assembled to get a coarse-grained image with fewer degrees of freedom than the original image we started with. The assembled coarse-gained image is again homogenized to find the overall homogenized elastic constants of the sample.

The idea of applying renormalization-based approach to obtain homogenized properties of rocks has attracted some attention in the past~\citep{Russel1985, King1989, King1996, Banerjee2004, Green2007,  Karim2010, Hanasoge2017, Hansen1997a, Wei2019}. However, a systematic analysis of the error incurred by the renormalization-based schemes is largely unavailable. The characterization and control of the associated error constitutes an essential step in order to confidently apply any approximation method. The present work attempts to characterize the error associated with the hierarchical homogenization method as applied to computing homogenized elastic moduli of rocks.

The renormalization idea of dividing and reassembling a domain is potentially powerful in reducing the computational expenses associated with determining the effective elastic properties of heterogeneous materials. However, to our best knowledge, surprisingly, there does not seem to be a substantial body of research devoted to this area. \citet{Banerjee2004} employed  the renormalization approach $-$ which they termed recursive cell method $-$ to determine the homogenized elastic properties of polymer bonded explosive with the aim of reducing computational cost. Their work was limited to relatively small two-dimensional problems; and employed artificial samples, consisting of random distribution of varying size of circular-shaped hard phases, instead of real microstructures of materials. They further assumed that the individual subdomains, after coarse graining, can be described by an isotropic elastic medium.  This assumption may be violated for small enough subdomains and add to coarse-graining error. Moreover, while noticing that the error introduced by the renormalization decreases with increasing size of the subdomains, the nature of the error has not been examined systematically. The present work seeks to relax some of the assumptions in the previous studies and analyze the error caused by the hierarchical homogenization approach in three-dimensional microstructures of real rocks.

The main contributions and findings of this work are summarized as follows. We apply the hierarchical homogenization method to two-phase heterogeneous microstructures obtained from micro-CT scanning of five different sandstone rock samples. The 3D image of size $N\times N\times N$ is partitioned into subimages of size $n\times n\times n$ ($n < N$) that are sufficiently larger than the correlation length of the microstructure. Each subimage is then homogenized into a fully anisotropic elastic medium, represented by a voxel in the coarse-grained image. The error of the hierarchical homogenization approach is obtained by comparing its prediction of the effective elastic moduli with the brute-force solution of the elasticity PDE on the original image. We show that the error in the hierarchically homogenized elastic moduli follows a power law scaling with exponent $-1$ with respect to $n$. The systematic error in the hierarchically homogenized elastic moduli is shown to reduce still further by introducing a padding of surrounding materials during the homogenization of smaller subimages, and the $\mathcal{O}(n^{-1})$ scaling is preserved even with padding. This power law scaling of error is then employed to further reduce the error in the predicted homogenized elastic constants by Richardson extrapolation.

The remaining part of this manuscript is organized as follows. Section~\ref{sec_rock_samples} briefly discusses the statistical properties of the rock samples studied in this work. Section~\ref{sec_hier_homo} presents the hierarchical homogenization method and discusses its main advantages in reducing the computational costs while controlling the error bounds. Section~\ref{sec_homo_fft_solver} presents the detailed steps involved in the homogenization of a single subimage. Section~\ref{sec_result} presents the results obtained from the application of the hierarchical homogenization method to compute the homogenized elastic constants, along with the associated error, of the five digital samples. Section~\ref{sec_conclusion} summarizes the main conclusions of this work.

\section{Microstructure of rock samples}
\label{sec_rock_samples}

\begin{figure}[ht!]
    \centering
    \includegraphics[width=0.8\textwidth]{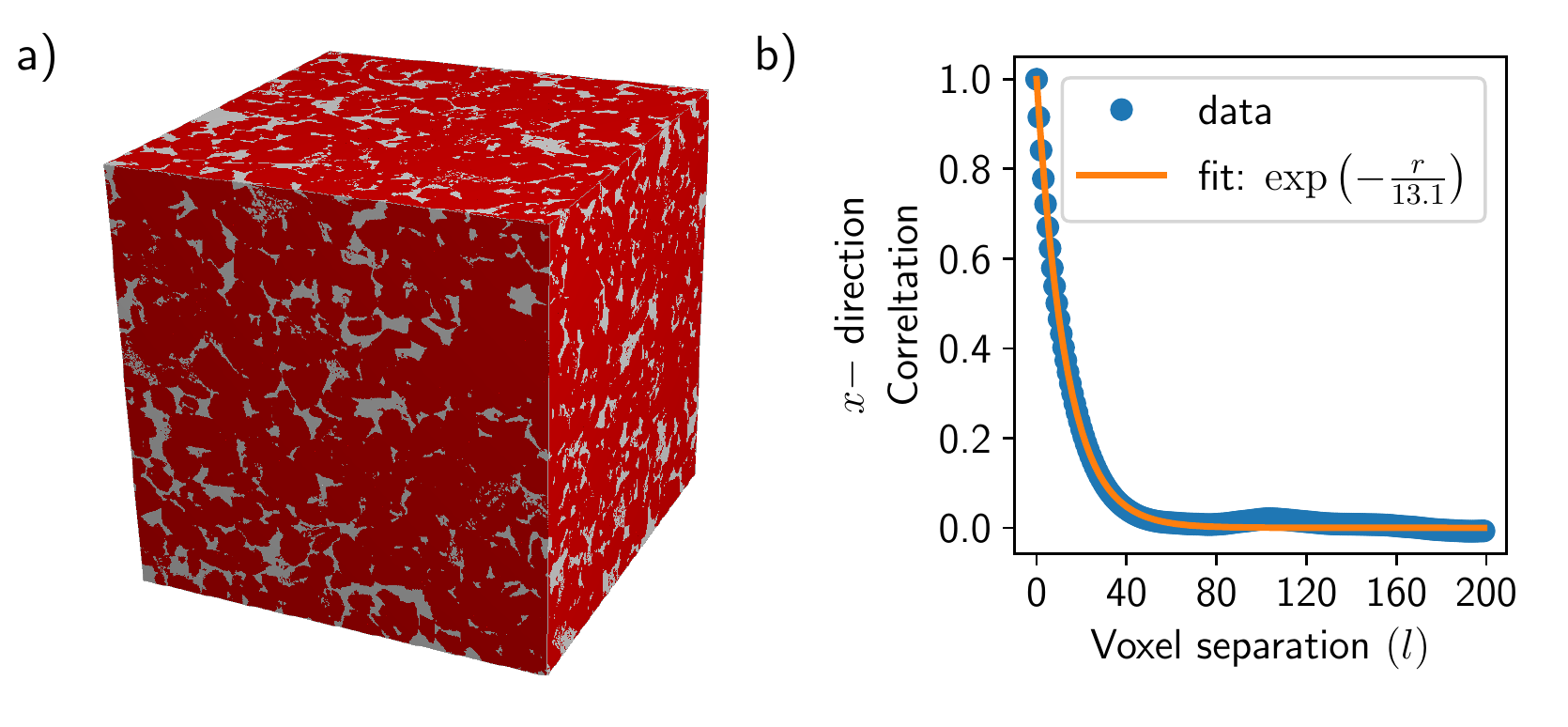}
    \caption{(a) Segmented images of rock samples B1. Red color denotes quartz mineral (phase 1) and white color represents pore (phase 2). (b) Two point correlation in sample B1 along the $x$-direction. Blue circles are the computed correlation function values, and orange curve is a fit to function $\exp(-l/\xi)$, where $\xi$ is the correlation length.}
    \label{fig_rock_sample_B1}
\end{figure}

The rock samples used in this work are the same as those in a previous study~\cite{Saxena2019}.  These include five microstructures obtained from three different sandstones: two (B1 and B2) are from the Berea Formation, subangular to subrounded Mississippian-age sandstone; two (FB1 and FB2) from the Fontainebleau Formation, subrounded to rounded Oligocene age sandstone; and one (CG) from the Castlegate Formation, subangular to subrounded Mesozoic sandstone. All of the rocks were imaged with a micro-CT scanner at the image resolution of approximately 2 $\mu m$ (the physical size of each voxel). The X-ray diffraction (XRD) analysis demonstrates that all samples mainly consist of quartz mineral along with trace amounts of feldspar, calcite and clay. For simplicity, we assume single mineralogy where all minerals were treated as quartz. The samples are then segmented into two phases of quartz (phase 1) and pore (phase 2) using the Otsu algorithm. Figure~\ref{fig_rock_sample_B1} shows the segmented images of sample B1 of size $N^3$ with $N = 900$. Red colored voxel corresponds to quartz mineral and white colored voxel to pore. The porosity of the five rock samples are given in Table~\ref{tab:rocks}. To facilitate comparisons across rock samples, we assume that the elastic constants of each phase do not vary across samples: minerals (phase 1) have been assigned the elastic properties of quartz, bulk modulus of $36$~GPa and shear modulus of $45$~GPa; and pores (phase 2) have both bulk and shear moduli equal to $0$ GPa~\citep{mavko2003rock} (for numerical stability, bulk and shear moduli of the pore phase are taken to be $10^{-5}$~GPa). For more information on acquisition and segmentation of micro-CT imaged rock samples, reader are directed to Ref.~\citep{Saxena2019}.

\begin{table}[ht!]
    \centering
    \caption{Statistical properties of the rock samples, including porosity and correlation lengths $\xi$ in $x$, $y$, and $z$ directions. }
    \begin{tabular}{ c  c c c c }
        \toprule
        Rock sample & porosity & $\xi_x$ & $\xi_y$ & $\xi_z$ \\
        \midrule
        B1          & 16.51\%  & 13.1    & 12.8    & 13.4    \\
        B2          & 19.63\%  & 14.2    & 15.3    & 14.5    \\
        CG          & 22.20\%  & 10.7    & 8.5     & 11.8    \\
        FB1         & 9.25\%   & 14.4    & 16.1    & 15.6    \\
        FB2         & 3.45\%   & 15.9    & 15.1    & 15.1    \\
        \bottomrule
    \end{tabular}
    \label{tab:rocks}
\end{table}

The correlation length is an important characteristic that measures the intrinsic length scale of a microstructure. To compute the correlation length, we first define a material function $M(i, j, k)$ at every voxel with indices $i, j, k$ from $1$ to $N$, such that

\begin{align}
    M(i,j,k) = \begin{cases}
        1 \quad \text{if voxel } (i,j,k) \text{ is occupied by mineral (phase 1)} \\
        0 \quad \text{if voxel } (i,j,k) \text{ is occupied by pore (phase 2)}.
    \end{cases}
\end{align}
The correlation between two points separated by $l$ voxels in the $x$-direction (corresponding to $i$ index) is given by
\begin{align}
    \text{Corr}_x(l) = \frac{\langle M(i,j,k) M(i+l, j, k) \rangle - \langle M(i,j,k) \rangle^2}{\langle M^2(i,j,k) \rangle - \langle M(i,j,k) \rangle^2 },
    \label{eq_correlation}
\end{align}
where $\langle \cdot \rangle$ represents the average operation over all possible values of indices $i,j$, and $k$. We can similarly define correlation functions in $y$- and $z$-directions.

Figure~\ref{fig_rock_sample_B1}(b) shows the two-point correlation function as a function of voxel separation $l$ in the $x$-direction for the rock sample B1. The blue circles are the values of the correlation function obtained using Equation~\eqref{eq_correlation}, and the orange curves are the best fit of the exponential function $\exp\left( -{l}/{\xi} \right)$, where $\xi$ is the correlation length. For sample B1 the correlation length in $x$-direction is $\xi = 13.1$.
The values of the correlation lengths $\xi$ for the five rock samples are given in Table~\ref{tab:rocks}.  The correlation lengths in all three directions are found to be similar, indicating that all the microstructures are isotropic to a good approximation. We observe that the correlation becomes very small ($<$ 0.05) at distances beyond $3 \xi$ (i.e. about 40 for rock sample B1). This information is helpful in deciding the representative volume element (RVE), i.e. the size of the subimages, for the hierarchical homogenization procedure.

\section{Hierarchical homogenization}
\label{sec_hier_homo}

This section lays out the hierarchical homogenization approach to determine the effective elastic constants of rocks. The goal of homogenization is to approximate a given random heterogeneous material by a homogeneous material which exhibits the same elastic response as the original heterogeneous material under some specific loading conditions. In other words, we seek a coarse-grained description of the heterogeneous material. Instead of homogenizing the whole domain (micro-CT scan image) at once by solving the elasticity PDE, we follow the renormalization based approach where the problem domain is scaled-down progressively by successive coarse-graining steps.

\begin{figure}[ht!]
    \centering
    \includegraphics[width=\textwidth]{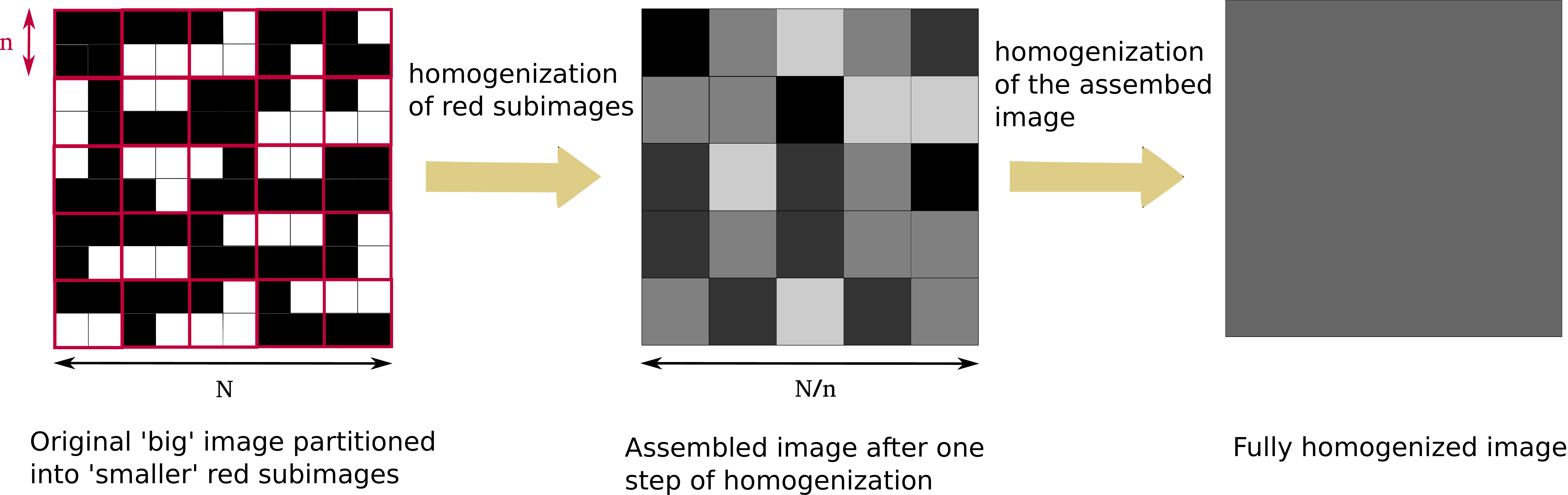}
    \caption{A schematic illustration of the two-step hierarchical homogenization scheme. A large image of segmented rock sample of size $N^3$ is partitioned into $(N/n)^3$ small subimages (shown in red color in the first figure) of size $n^3$. Each red subimage is homogenized by solving the local elasticity problem and assembled into an intermediate image of reduced size $(N/n)^3$. The assembled intermediate image is finally homogenized to obtain the homogenized properties of the original image. Black and white color in the left panel denotes the two individual phases of the rock. The gray level in the middle panel represents the homogenized value of the relevant properties of the corresponding subimage of size $n^3$.}
    \label{fig_hh_schematic}
\end{figure}

As schematically depicted in Figure~\ref{fig_hh_schematic}, we start with a large 3D image of size $N\times N\times N$. This image is then partitioned into $\left(N/n\right)^3$ smaller subimages of size $n$ which are shown in the red color in the first subfigure of Figure~\ref{fig_hh_schematic}. Each subimage consists of two phases that are assumed to be isotropic elastic having different elastic constants. We then seek to replace this small subimage by a single voxel, i.e. $n^3$ voxels are decimated into just one voxel that summarizes the (anisotropic) elastic properties of the corresponding subimage. To accomplish this decimation, we solve the local elasticity problem for all $(N/n)^3$ smaller subimages separately under periodic boundary conditions and determine their corresponding average elastic constants, as described in Section~\ref{sec_homo_fft_solver}. In the next step, the average elastic constants of the subimages are used to assemble a coarse-grained version of the original image. The coarse-grained image now has $(N/n)^3$ voxels (see middle panel of Figure~\ref{fig_hh_schematic}) compared with the $N^3$ voxels of the original image. The assembled coarse-grained image retains the low-frequency information of the original image, while the local high-frequency components are captured in the properties of the small subimages. In the renormalization literature, this coarse-graining/homogenization/decimation is often repeated multiple times until a fixed point is achieved~\cite{Wilson1983}. However, in this work, we carry out just two steps of coarse-graining, as depicted in Figure~\ref{fig_hh_schematic}. The assembled coarse-grained image is homogenized one more time to find the effective average elastic constants of the rock sample.

\begin{figure}[ht!]
    \centering
    \includegraphics[width=0.9\textwidth]{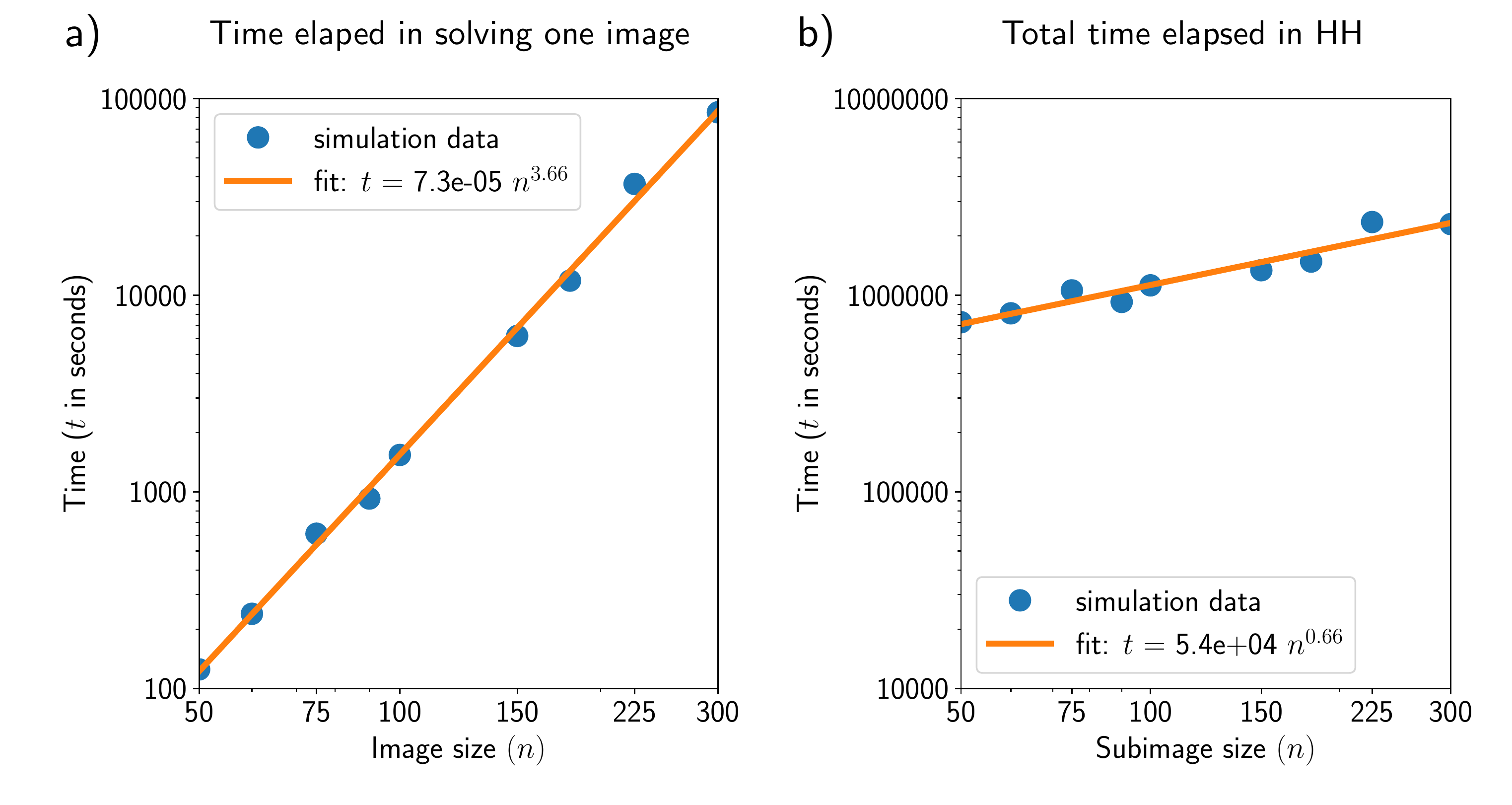}
    \caption{ Log-log plot of (a) time to solve a single image as a function of image size, (b) total time elapsed during the hierarchical homogenization (HH) of a big image of size $N = 900$ as a function of subimage size $n$. The calculations are performed using an FFT-based elasticity solver. Blue dots denote the measured computational time, and the orange line is the best fit of the data to a power law. }
    \label{fig_time_scaling}
\end{figure}

We now discuss the numerical advantages associated with the hierarchical homogenization scheme. Figure~\ref{fig_time_scaling}(a) shows the time elapsed for solving an elasticity problem on an image with $n^3$ voxels. Blue dots are the measured computational time data, and the orange line is obtained by fitting the data to a power law. The computation time scales as $\mathcal{O}\left(n^{3.66}\right)$ which is super-linear with respect to the number of voxels ($n^3$) in the image. We now consider a large image containing $N^3$ voxels. Instead of solving the elasticity PDE on this image, we divide it into $(N/n)^3$ smaller subimages of size $n^3$. Then the time $t$ spent in solving $(N/n)^3$ subimages scales as
\begin{align}
    t \sim \left(\frac{N}{n}\right)^3 \mathcal{O}\left(n^{3.66}\right) \sim \mathcal{O}\left( n^{0.66} \right).
\end{align}

Neglecting the time it takes to solve the elasticity problem on the coarse-grained image, the total computation time to solve a big image with $N^3$ voxels thorough hierarchical homogenization (essentially solving $(N/n)^3$ subimages) scales with the size of the subimage as $\mathcal{O}(n^{0.66})$ as shown in Figure~\ref{fig_time_scaling}(b). The smaller the subimage, the shorter the computation time. The computation time shown in Figure~\ref{fig_time_scaling} is for a specific set of elastic constants of the two phases, and the exact computation time will be highly dependent on the contrast between the properties of the underlying constituent phases and the implementation details of the algorithm. The gain in computation time, nevertheless, will result for any computational algorithm with a time complexity that is super linear with the number of voxels. We further note that aside from the gain in computation time, the hierarchical homogenization scheme also reduces the memory requirement that would be prohibitive for the full scale simulation. For example, the computational resources available to us limit the largest image that can be solved by brute-force to $1000^3$ voxels due to memory constraints.

\section{Homogenization of a single subimage}
\label{sec_homo_fft_solver}

This section discusses the two steps involved in determining the average elastic constants of the subimages: (1) extracting the subimage from the large image; (2) averaging the equilibrium stress and strain fields inside the subimage.

\begin{figure}
    \centering
    \includegraphics[width=\textwidth]{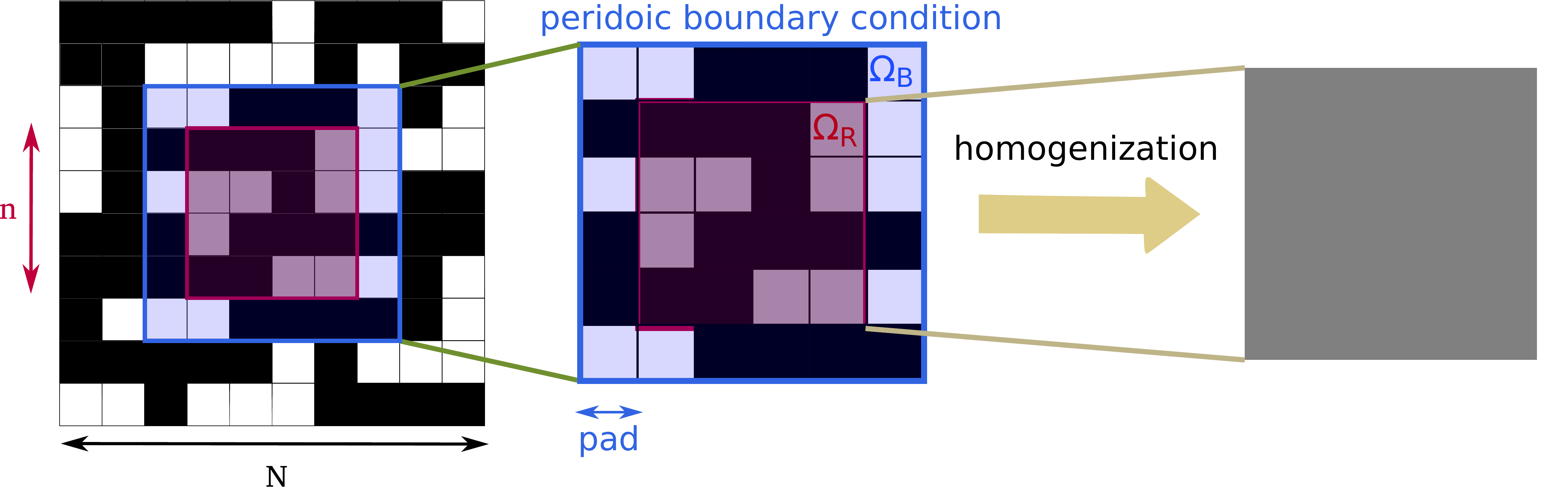}
    \caption{Schematic diagram illustrating the extraction and homogenization of the subimages. The goal is to find the average elastic constants of the subimage shown in the color red occupying the region $\Omega_{\rm R}$. In order to mitigate the boundary effects and mimic the original environment experienced by the red subimage in the original large image, we add extra padding of the surrounding materials to form the blue region $\Omega_{\rm B} \supseteq \Omega_{\rm R}$. We then solve the local elasticity problem in the extracted blue image $\Omega_{\rm B}$ subject to periodic boundary conditions. We finally average stress and strain fields inside the red region $\Omega_{\rm R}$ to find the average anisotropic elastic constant of the red subimage.
    }
    \label{fig_subimage_padding}
\end{figure}

Figure~\ref{fig_subimage_padding} illustrates the extraction step for a single subimage occupying the domain $\Omega_{\rm R}$. In order to compute the effective elastic constants of the region $\Omega_{\rm R}$, it is necessary to solve the elasticity PDE on this domain subjected to appropriate boundary conditions. For example, it is common practice to apply periodic boundary conditions (PBC) when an FFT-based solver is used. However, the application of PBC (or any other types of boundary conditions) to region $\Omega_{\rm R}$ introduces an error because the voxels in $\Omega_{\rm R}$ are subjected to a different environment compared to that in the original image. Because such an error is introduced at the boundary of every subimage, we expect the impact of this error on the final homogenized elastic constants to decrease with the size of the subimage dimension $n$.

To control this error due to the subimage boundary conditions, we choose to extract a larger domain of voxels, $\Omega_{\rm B}$, which includes the domain $\Omega_{\rm R}$ and a padding region, as shown in Figure~\ref{fig_subimage_padding}.  The inclusion of this padding region enables, to an extent, the voxels in $\Omega_{\rm R}$ to experience a similar environment to that in the original image. According to Saint-Venant's principle, given a sufficiently thick padding region, the details of the boundary conditions will not matter to the stress-strain solution in domain $\Omega_{\rm R}$, as long as the net force and moment transmitted through each boundary face stays the same. Therefore, we expect the error due to subimage boundary conditions to decay rapidly with the padding region thickness.

In this work, we solve the elasticity PDE on domain $\Omega_{\rm B}$ subject to PBC.  The resulting stress and strain fields are then used to compute the average stress and strain values in the domain of interest $\Omega_{\rm R}$, and its effective elastic constants. Thicker padding region would lead to more accurate results, but at the expense of higher computational cost. Hence, the choice of the padding thickness is the result of a trade-off between accuracy and computational cost.

We now solve the elasticity PDE in the domain $\Omega_{\rm B}$ subjected to PBC. The elastic response of the material at a point $\bs{x} \in \Omega_{\rm B}$ is prescribed by a position dependent isotropic elastic tensor $\mathbb{C}_{ijkl}(\bs{x}) = \left(K(\bs{x}) - \frac{2}{3} \mu(\bs{x})\right) \delta_{ij} \delta_{kl} + \mu(\bs{x}) \left(\delta_{ik} \delta_{jl} + \delta_{il} \delta_{jk}\right)$, where $K(\bs{x})$ and $\mu(\bs{x})$ are bulk and shear moduli at point $\bs{x}$. Thus, the local stress $\sigma_{ij}(\bs{x})$ relates to the local strain $\varepsilon_{ij}(\bs{x})$ via
\begin{align}
    \sigma_{ij}(\bs{x}) = \mathbb{C}_{ijkl}(\bs{x}) \varepsilon_{kl}(\bs{x}) = K(\bs{x}) \varepsilon_{kk}(\bs{x}) \delta_{ij} + 2\mu(\bs{x}) \left(\varepsilon_{ij}(\bs{x}) - \frac{1}{3}\varepsilon_{kk}(\bs{x})\delta_{ij}\right),
\end{align}
where Einstein's notation is assumed, in which repeated indices are summed over from 1 to 3. We now assume that the domain $\Omega_{\rm B}$ is subjected to an average strain $E_{ij}$ so that the local strain field can be decomposed into two parts, i.e.
\begin{align}
    \varepsilon_{ij}(\bs{x}) = E_{ij} + \varepsilon^*_{ij}(\bs{x}), \qquad \int_{\Omega_{\rm B}} \varepsilon^*_{ij}(\bs{x}) = 0,
\end{align}
where $E_{ij}$ corresponds to a homogenized applied strain, and $\varepsilon^*_{ij}(\bs{x})$ is the $\Omega_{\rm B}$-periodic fluctuation due to the heterogeneity of the microstructure. The fluctuation in the strain field must satisfy both the compatibility and equilibrium conditions, i.e.
\begin{equation}
    \begin{aligned}
        \varepsilon^*_{ij} (\bs{x})                                 = \frac{1}{2}(u^*_{i,j}(\bs{x}) + u^*_{j,i}(\bs{x}))  \quad \forall \bs{x} \in \Omega_{\rm B},  \qquad \text{(compatibility condition)} \\
        \sigma_{ij,j}(\bs{x}
        )  = 0                                                   \quad \forall \bs{x} \in \Omega_{\rm B}, \qquad \text{(equilibrium condition)}
    \end{aligned}
    \label{eq_equilibrium_compatibility}
\end{equation}
where $ u^*_i(\bs{x})$  is an $\Omega_{\rm B}$-periodic displacement field, and $_{,j} \equiv \partial/\partial x_j$.

Equation ~\eqref{eq_equilibrium_compatibility} can be solved by using various numerical methods, such as finite element or fast fourier transform (FFT) based solvers, to obtain equilibrium stress $\sigma_{ij}(\bs{x})$ and strain $\varepsilon_{ij}(\bs{x})$ fields in region $\Omega_{\rm B}$.
Here we employ an FFT-based method whose basic formulation is presented in~\ref{app_fft_solver}. To compute the homogenized elastic constants in domain $\Omega_{\rm R}$ (excluding the padding), we take the average of the obtained stress and strain fields over this domain.
\begin{equation}
    \begin{aligned}
        \langle \sigma \rangle_{{\rm R},ij}      & = \frac{1}{| \Omega_{\rm R} |} \int_{\Omega_{\rm R}} d\bs{x} \, \sigma_{ij}(\bs{x}),                              \\
        \langle \varepsilon \rangle_{{\rm R},ij} & = \frac{1}{| \Omega_{\rm R} |} \int_{\Omega_{\rm R}} d\bs{x} \left( E_{ij} +  \varepsilon^*_{ij}(\bs{x}) \right).
    \end{aligned}
\end{equation}
By definition, the homogenized elastic constant $\mathbb{C}^{\rm homo}_{{\rm R}, ijkl}$ of the subimage satisfies the following condition
\begin{equation}
    \langle \sigma \rangle_{{\rm R},ij}  = \mathbb{C}^{\rm homo}_{{\rm R},ijkl} \, \langle \varepsilon \rangle_{{\rm R},kl}.
\end{equation}

To extract the homogenized elastic constants, it is more convenient to use the Voigt notation, in which the stress and strain tensors are represented as $6\times 1$ column vectors ($\uvec{\langle \sigma \rangle}_{\rm R}$ and $\uvec{\langle \varepsilon \rangle}_{\rm R}$) and the elastic stiffness tensor as a $6\times 6$ matrix ($\mat{C}_{\rm R}^{\rm homo}$).
In the Voigt notation, the homogenized stress-strain relation in domain $\Omega_{\rm R}$ is,
\begin{align}
    \uvec{\langle \sigma \rangle}_{\rm R}  = \mat{C}_{\rm R}^{\rm homo} \, \uvec{\langle\varepsilon\rangle }_{\rm R} \, .
    \label{eq_homo_stiffness_voigt}
\end{align}
In order to compute homogenized stiffness matrix $\mat{C}_{\rm R}^{\rm homo}$, six independent macroscopic strain fields $(E_{ij})$ are imposed and the elasticity PDE is then solved numerically to give six pairs of $\uvec{\langle \sigma \rangle}_{\rm R}$ and $\uvec{\langle\varepsilon\rangle }_{\rm R}$ vectors. We concatenate the six stress vectors to form a $6\times 6$ matrix $\mat{\Sigma}_{\rm R}$, and concatenate the six strain vectors to form a $6\times 6$ matrix $\mat{\mathcal{E}}_{\rm R}$. These matrices still satisfy the stress-strain relation,
\begin{align}
    \mat{\Sigma}_{\rm R} =
    \mat{C}^{\rm homo}_{\rm R} \, \mat{\mathcal{E}} _{\rm R} \, .
\end{align}
Therefore, the homogenized elastic stiffness tensor can be obtained by
\begin{align}
    \mat{C}^{\rm homo}_{\rm R} = \mat{\Sigma} _{\rm R}~ \mat{\mathcal{E}}^{-1}_{\rm R} \, .
\end{align}
In general, the resulting homogenized stiffness matrix $\mat{C}^{\rm homo}_{\rm R}$ is fully anisotropic.  Hence, in the second step of the hierarchical homogenization approach, the coarse-grained image consists of voxels each representing an anisotropic elastic medium, even if the voxels in the original image represents isotropic elastic medium.

In this work we use ElastoDict~\citep{Elastodict, Kabel2013, Kabel2016} module of GeoDict software to solve the isotropic elastic problem for small subimages; and to solve fully anisotropic elasticity problem in the intermediate assembled image, we modify the python code, freely available on webpage~\citep{Geus}, of which basic algorithms are detailed in Refs.~\citep{DeGeus2017a, Zeman2017}.

\section{Results}
\label{sec_result}

In this section, we discuss results obtained from the hierarchical homogenization scheme of five different rocks of size $N^3$ with $N = 900$.  We pay special attention on the influence of subimage size and the padding thickness on the error in the overall homogenized elastic properties. The large image is partitioned into subimages of size $n^3$ with $n = 75$, $90$, $100$, $150$, $180$,  $225$, $300$. Padding thicknesses of 0, 10, and 20 are used for each partition scheme.

\begin{figure}[ht!]
    \centering
    \includegraphics[width=\textwidth]{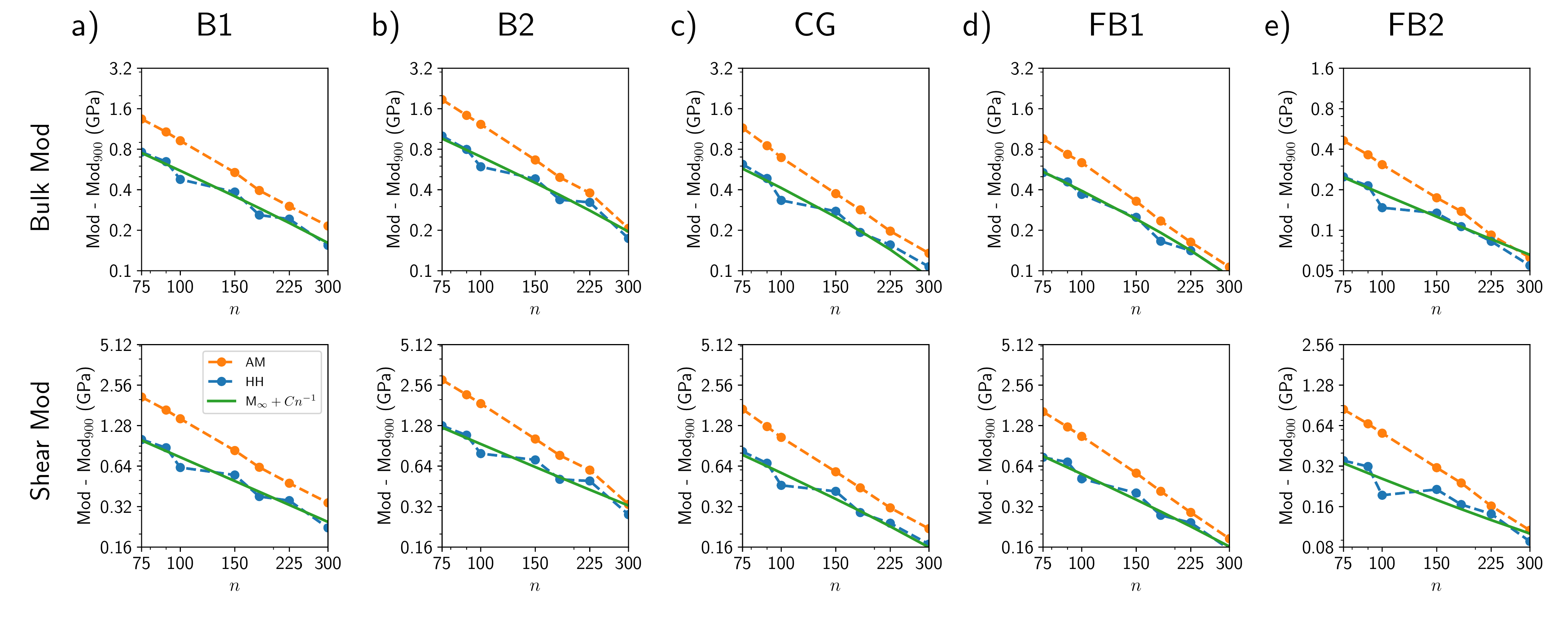}
    \caption{Effect of subimage size on the hierarchically homogenized elastic constants for zero padding case for the five rock samples of size $N = 900$: (a) B1, (b) B2, (c) CG, (d) FB1, and (e) FB2. In each subfigure, corresponding to a specific rock sample, the variation of errors in the homogenized elastic moduli obtained from the hierarchical homogenization are plotted on log-log scale as a function of the subimage size $n$. The error is calculated by subtracting the corresponding homogenized elastic moduli of the big image of size $n=900$ that is obtained by solving the local elasticity problem in the big image. The first panel in each subfigure shows the result for the bulk modulus and the second panel for the shear modulus. The orange plots show the arithmetic mean (AM) of the elastic moduli of the small images composing the big image; the blue plots show the result obtained from the hierarchical homogenization (HH); and the green line is the fit of the hierarchical homogenization values to the function $M_{\infty} + C n^{-1}$, with $M_\infty$ and $C$ being the two fitting parameters.}
    \label{fig_result_900}
\end{figure}

We first discuss the results for zero padding.  For comparison purposes, we also compute the simple arithmetic averages of the elastic moduli (shear and bulk moduli) of all the subimages of a given partition of the original image which is precisely the Voigt averaging. The error of each homogenization scheme is computed by comparing with the direct solution of the elasticity PDE on the original image of size $N^3$. Figure~\ref{fig_result_900} presents the error in the homogenized elastic moduli as obtained from the hierarchical homogenization scheme and from simple arithmetic averages as a function $n$ for all five rock samples. The modulus values are also listed in Table~\ref{table_result_hh_900_real_rock}. We note that the errors across all cases are positive, indicating a systematic error caused by the periodic boundary condition used for the homogenization of the subimages~\citep{Huet1990, Hazanov1994, Kanit2003a}.

Figure~\ref{fig_result_900} shows that the error in the homogenized elastic moduli decreases with increase in subimage size $n$ for both hierarchical homogenization and arithmetic mean values. The monotonic decrease in the error for the hierarchical homogenization indicates that the rock microstructure is fairly homogeneous at the length scale considered here ($n \ge 75$). This is consistent with the correlation lengths given in Table~\ref{tab:rocks}. The error for the arithmetic mean is larger than the hierarchical homogenization value for the same subimage size. Therefore, assembling the homogenized subimages into voxels in a coarse-grained image provides a more accurate description of the overall elastic property than simple averages of the elastic stiffness of the subimages.

\begin{table}[ht!]
    \centering
    \caption{ Homogenized elastic moduli (bulk and shear moduli) of the five rock images of size $N=900$ as a function of small image size $n$ into which the big image is partitioned. The values are obtained by the hierarchical homogenization (HH) method and by taking arithmetic mean (AM) of the homogenized elastic moduli of the small images. }
    \begin{tabular}{l c  c c c c c c c c  }
        \toprule
        \multicolumn{3}{c}{\diagbox{Rock type}{$n$}} & 75                               & 90 & 100   & 150   & 180   & 225   & 300                   \\
        \midrule
        \multirow{4}{*}{B1}                          & \multirow{2}{*}{Bulk mod (GPa)}  & HH & 22.06 & 21.95 & 21.78 & 21.69 & 21.56 & 21.54 & 21.46 \\
                                                     &                                  & AM & 22.64 & 22.38 & 22.23 & 21.84 & 21.70 & 21.60 & 21.52 \\
        \cline{ 3-10 }
                                                     & \multirow{2}{*}{Shear mod (GPa)} & HH & 24.51 & 24.38 & 24.13 & 24.05 & 23.89 & 23.86 & 23.73 \\
                                                     &                                  & AM & 25.59 & 25.18 & 24.94 & 24.34 & 24.13 & 23.98 & 23.85 \\
        \midrule
        \multirow{4}{*}{B2}                          & \multirow{2}{*}{Bulk mod (GPa)}  & HH & 19.83 & 19.62 & 19.42 & 19.31 & 19.16 & 19.15 & 19.00 \\
                                                     &                                  & AM & 20.70 & 20.25 & 20.05 & 19.49 & 19.32 & 19.20 & 19.03 \\
        \cline{ 3-10 }
                                                     & \multirow{2}{*}{Shear mod (GPa)} & HH & 21.58 & 21.39 & 21.09 & 21.01 & 20.81 & 20.80 & 20.58 \\
                                                     &                                  & AM & 23.10 & 22.47 & 22.17 & 21.32 & 21.07 & 20.90 & 20.63 \\
        \midrule
        \multirow{4}{*}{CG}                          & \multirow{2}{*}{Bulk mod (GPa)}  & HH & 18.39 & 18.26 & 18.11 & 18.05 & 17.97 & 17.93 & 17.88 \\
                                                     &                                  & AM & 18.92 & 18.62 & 18.47 & 18.15 & 18.06 & 17.97 & 17.91 \\
        \cline{ 3-10 }
                                                     & \multirow{2}{*}{Shear mod (GPa)} & HH & 19.78 & 19.63 & 19.42 & 19.37 & 19.25 & 19.20 & 19.13 \\
                                                     &                                  & AM & 20.65 & 20.22 & 20.00 & 19.54 & 19.40 & 19.27 & 19.18 \\
        \midrule
        \multirow{4}{*}{FB1}                         & \multirow{2}{*}{Bulk mod (GPa)}  & HH & 28.16 & 28.08 & 27.99 & 27.87 & 27.79 & 27.77 & 27.72 \\
                                                     &                                  & AM & 28.58 & 28.36 & 28.26 & 27.95 & 27.86 & 27.79 & 27.73 \\
        \cline{ 3-10 }
                                                     & \multirow{2}{*}{Shear mod (GPa)} & HH & 33.12 & 33.06 & 32.89 & 32.78 & 32.65 & 33.61 & 32.53 \\
                                                     &                                  & AM & 33.99 & 33.62 & 33.44 & 32.94 & 32.79 & 32.66 & 32.56 \\
        \midrule
        \multirow{4}{*}{FB2}                         & \multirow{2}{*}{Bulk mod (GPa)}  & HH & 33.09 & 33.06 & 32.99 & 32.98 & 32.95 & 32.93 & 32.90 \\
                                                     &                                  & AM & 33.31 & 33.21 & 33.15 & 33.02 & 32.98 & 32.94 & 32.91 \\
        \cline{ 3-10 }
                                                     & \multirow{2}{*}{Shear mod (GPa)} & HH & 40.46 & 40.42 & 40.30 & 40.32 & 40.27 & 40.24 & 40.19 \\
                                                     &                                  & AM & 40.95 & 40.76 & 40.67 & 40.41 & 40.34 & 40.27 & 40.21 \\
        \bottomrule
    \end{tabular}
    \label{table_result_hh_900_real_rock}
\end{table}

For the hierarchical homogenization, we now quantify the convergence rate of the error with respect to the size of the small image $n$.  As shown in Figure~\ref{fig_result_900}, errors in both elastic bulk and shear moduli for the hierarchical homogenization case (blue dots) decrease with increasing $n$ across all five rock samples. Furthermore, on the log-log scale, the variation in error is almost linear, suggesting a power-law relationship between the error and $n$. The fact that error is found by subtracting a fixed value (elastic moduli of image of size $N=900$) from the hierarchically homogenized elastic moduli means that the hierarchically homogenized elastic moduli itself can be well-fitted to a power-law with exponent $-1$ with an additional additive parameter $M_{\infty}$, i.e.
\begin{align}
    M^{\rm HH}(n) = M_{\infty} + C \, n^{-1},
    \label{eq_fit_power_law}
\end{align}
where $M^{\rm HH}(n)$ stands for both bulk and shear moduli computed from the hierarchical homogenization method using subimages of size $n^3$; and $M_{\infty}$ and $C$ are the two fitting parameters that, respectively, represent the asymptotic modulus value and the rate of convergence. The reasoning behind fixing the exponent value to $-1$ is that the major source of error arises from the interface between the two neighboring small subimages. The homogenized subimages capture the bulk microstructure features present within the image. However, the microstructure features located across the interface of the two neighboring subimages get smeared out in the first step of the hierarchical homogenization method, and results in the error in the final homogenized elastic moduli. Therefore, the error in the final elastic moduli scales as the ratio between the area and volume of the subimages which is $n^{-1}$.

\begin{table}[ht!]
    \centering
    \caption{Homogenized properties of the five different rock samples. $M_{900}$ denotes the elastic modulus obtained by directly solving the big image of size $N=900$. $M_{\infty}$ is the asymptotic value of the corresponding elastic modulus that is obtained by fitting the hierarchical homogenization result to the power law given in Equation~\eqref{eq_fit_power_law}. The value of $M_\infty$ is a good approximation for the homogenized elastic modulus of a big rock sample.}
    \begin{tabular}{l  c   c  c  c   c  c c  }
        \toprule
        \multirow{2}{*}{Rock type} & \multirow{2}{*}{Porosity(\%)} & \multicolumn{3}{c}{Bulk modulus (GPa)} & \multicolumn{3}{ c }{Shear modulus (GPa)}                                            \\
        \cline{3-8}
                                   &                               & $M_{900}$                              & $M_{\infty}$                              & $C$   & $M_{900}$ & $M_{\infty}$ & $C$   \\
        \midrule
        B1                         & 16.51                         & 21.30                                  & 21.26                                     & 58.95 & 23.51     & 23.50        & 74.89 \\
        B2                         & 19.63                         & 18.82                                  & 18.77                                     & 76.44 & 20.30     & 20.32        & 91.28 \\
        CG                         & 22.20                         & 17.77                                  & 17.71                                     & 48.38 & 18.96     & 18.91        & 61.60 \\
        FB1                        & 9.25                          & 27.63                                  & 27.57                                     & 45.27 & 32.37     & 32.33        & 59.65 \\
        FB2                        & 3.45                          & 32.85                                  & 32.85                                     & 18.13 & 40.10     & 40.13        & 23.53 \\
        \bottomrule
    \end{tabular}
    \label{table_fit_results_900}
\end{table}

In Figure~\ref{fig_result_900}, the results of the fitting procedure are plotted in green. The qualities of the power-law scaling with fix exponent value of -1 are found to be good for both elastic moduli across all five rock samples. The values of these fitting parameters, $M_\infty$ and $C$, along with the elastic moduli $M_{900}$ of the image size 900 for all five rock samples are given in Table~\ref{table_fit_results_900}. The values of $M_\infty$ represents the asymptotic value of the corresponding elastic moduli, and  approximates really well the homogenized elastic moduli of a large image. Thus, the hierarchical homogenization scheme can be employed to compute the homogenized elastic moduli of a big rock image without ever solving the local elasticity problem in the big original image, thus gaining in time and avoiding the memory constraints encountered routinely in the application of computational methods to large domains.

\begin{figure}[ht!]
    \centering
    \includegraphics[width=\textwidth]{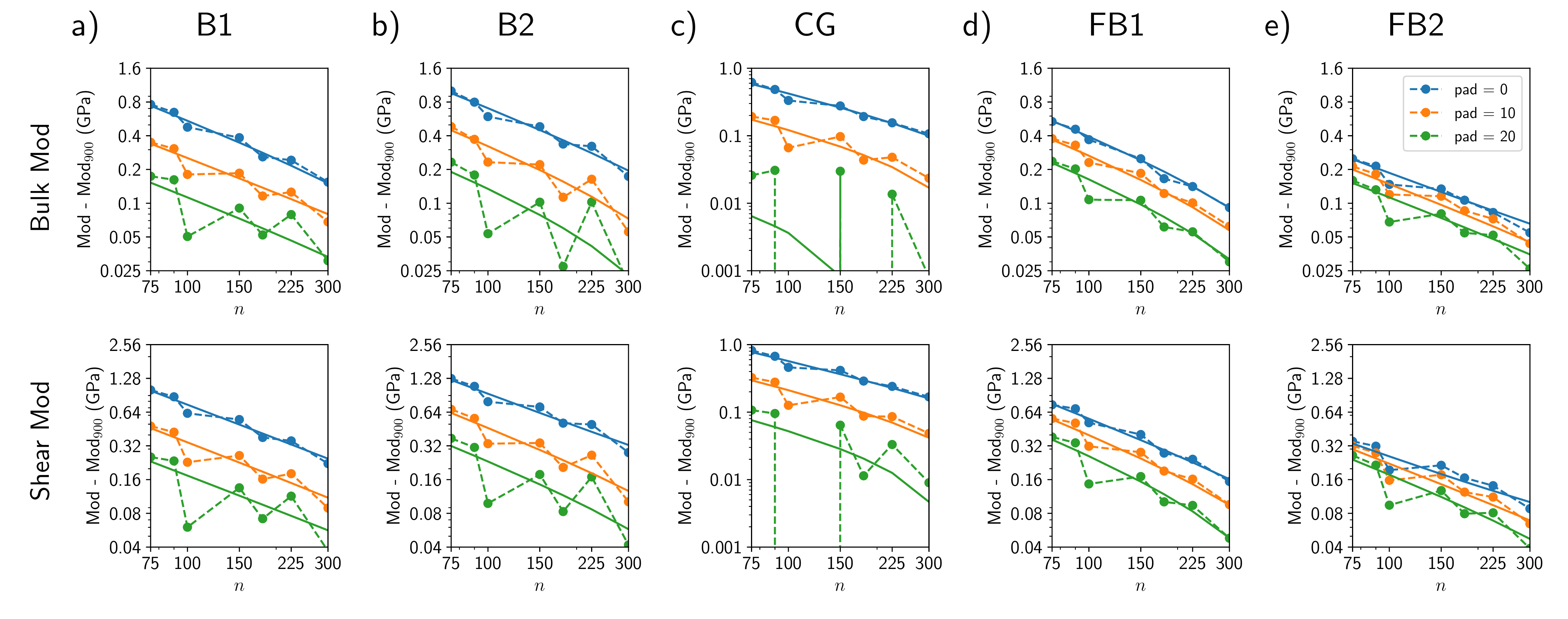}
    \caption{Effect of the padding thickness on the hierarchically homogenized elastic constants for different padding thickness for the five rock samples of size $N = 900$: (a) B1, (b) B2, (c) CG, (d) FB1, and (e) FB2. In each subfigure, corresponding to a specific rock sample, the variation of errors in the homogenized elastic moduli obtained from the hierarchical homogenization are plotted on log-log scale as a function of the subimage size $n$ for the three values of padding thickness: 0, 10, 20. The error is calculated by subtracting the corresponding homogenized elastic moduli of the big image of size $n=900$ that is obtained by solving the local elasticity problem in the big image. The first panel in each subfigure shows the result for the bulk modulus and the second panel for the shear modulus. The three colors correspond to the three values of the padding thickness: blue for 0, orange for 10, and green for 20. The dashed lines with circle markers show the simulation result and the solid curve is the fit of the hierarchical homogenization values to the function $M_{\infty} + C n^{-1}$, with $M_\infty$ and $C$ being the two fitting parameters. Across all the five rocks, the error decreases with increasing padding thickness for a fixed subimage size.}
    \label{fig_result_900_pad}
\end{figure}

Figure~\ref{fig_result_900_pad} demonstrates the effect of including padding materials surrounding the  subimages on the hierarchically homogenized elastic constants (bulk and shear moduli) of the corresponding big image. Three padding thickness, 0, 10, 20, are used, and the resulting error are plotted in blue, orange and green color in Figure~\ref{fig_result_900_pad}. The error in the hierarchically homogenized elastic moduli of the big image consistently decreases with increasing padding thickness across all the five rock samples. This means that the same error can be achieved using smaller subimages when some padding materials are included. For instance, for the bulk modulus of B1 rock sample, the error incurred by subimage size of 300 with zero padding is the same as that obtained from subimage size of 175 with padding thickness of 10. As discussed in Section~\ref{sec_hier_homo} and shown in Figure~\ref{fig_time_scaling}, the overall computational time will be shorter for hierarchical homogenization using subimages of size 175 with padding thickness of 10 than that for subimages of size 300 with zero padding.

The hierarchically homogenized elastic constants are further observed to follow the power law $M_\infty + C n^{-1}$, with exponent value of -1, for all three padding thickness values. The curves obtained from fitting the power law to the simulation data are plotted in solid lines in Figure~\ref{fig_result_900_pad}. Moreover, the fitting parameters ($M_\infty$ and $C$) are presented in Table~\ref{table_fit_pad} of the five rock samples, each with three padding thickness values. The parameter $M_\infty - $ representing the asymptotic hierarchically homogenized elastic constants $- $ is almost unchanged with padding thickness for both bulk and shear modulus. The value of $C$, however, generally decreases with increasing padding thickness which indicates a faster convergence to the asymptotic value. There is a relatively little change in the value of $C$ for FB2 rock sample. This is presumably caused by the dominance of one phase (phase 1) in the composition of the material.

\begin{table}
    \centering
    \caption{The result of the fitting of the power law expressions to the hierarchically homogenized elastic moduli (bulk and shear moduli) of the five rock samples with varying padding thicknesses. $M_\infty$ is the asymptotic value of the corresponding elastic modulus and $C$ controls the convergence rate of the power law. }
    \begin{tabular}{c c c c c c}
        \toprule
        Rock type            & Padding thickness & \multicolumn{2}{c}{Bulk Modulus (GPa)} & \multicolumn{2}{c}{Shear Modulus (GPa)}                      \\
        \cline{3-6}
                             &                   & $M_\infty$                             & $C$                                     & $M_\infty$ & $C$   \\
        \midrule
        \multirow{3}{*}{B1}  & 0                 & 60.40                                  & 78.71                                   & 26.67      & 18.64 \\
                             & 10                & 60.40                                  & 59.79                                   & 26.67      & 14.31 \\
                             & 20                & 60.42                                  & 43.31                                   & 26.67      & 9.74  \\
        \midrule
        \multirow{3}{*}{B2}  & 0                 & 55.21                                  & 99.77                                   & 24.86      & 24.18 \\
                             & 10                & 55.19                                  & 81.32                                   & 24.85      & 20.53 \\
                             & 20                & 55.22                                  & 62.80                                   & 24.85      & 15.39 \\
        \midrule
        \multirow{3}{*}{CG}  & 0                 & 51.77                                  & 75.83                                   & 23.60      & 20.73 \\
                             & 10                & 51.77                                  & 55.88                                   & 23.58      & 16.14 \\
                             & 20                & 51.80                                  & 36.83                                   & 23.59      & 10.13 \\
        \midrule
        \multirow{3}{*}{FB1} & 0                 & 74.51                                  & 81.89                                   & 31.23      & 14.22 \\
                             & 10                & 74.51                                  & 75.18                                   & 31.23      & 14.15 \\
                             & 20                & 74.53                                  & 63.42                                   & 31.22      & 11.55 \\
        \midrule
        \multirow{3}{*}{FB2} & 0                 & 88.49                                  & 43.24                                   & 35.16      & 5.58  \\
                             & 10                & 88.50                                  & 42.75                                   & 35.16      & 6.22  \\
                             & 20                & 88.49                                  & 38.50                                   & 35.17      & 5.61  \\
        \bottomrule
    \end{tabular}
    \label{table_fit_pad}
\end{table}

\section{Discussion and conclusion}
\label{sec_conclusion}


In this work, we apply a renormalization inspired method, the hierarchical homogenization method, to determine the average elastic constants of a random heterogeneous materials. The microstructures obtained from micro-CT scan of five real sandstone rock samples $-$ two from Berea Formation, one from Castlegate Formation, two from Fontainebleau Formation $-$ are used as the heterogeneous materials. The micro-CT scanned three-dimensional images are segmented into two phases (pore and mineral) are assigned to two different isotropic elastic materials.

In the hierarchical homogenization scheme, a large image is partitioned into multiple disjoint smaller subimages. The subimages are homogenized separately by solving local elasticity problems. The homogenized subimages are then assembled and homogenized to calculate the average elastic constants of the original big image. The hierarchical homogenization thus obviates the need to solve the local elasticity problem in the big domain and, thus, alleviating the prohibitive requirement of computational resources in terms of computational time and memory. This is particularly advantageous for cases where capturing the small length-scale features of the microstructure is important to produce accurate elastic properties. For instance, recent work on digital rock physics~\citep{Saxena2019} indicates that the insufficient resolution of the micro-CT scans leads to a systematic error in the computed average elastic constants. Application of high-resolution imaging modality would dramatically increase the number of voxels for a given representative volume and, therefore, render solving the local elasticity problem in the full image impractical. The hierarchical homogenization approach makes it feasible to predict the effective elastic moduli from these large images.
%

\begin{figure}[ht!]
    \centering
    \includegraphics[width=\textwidth]{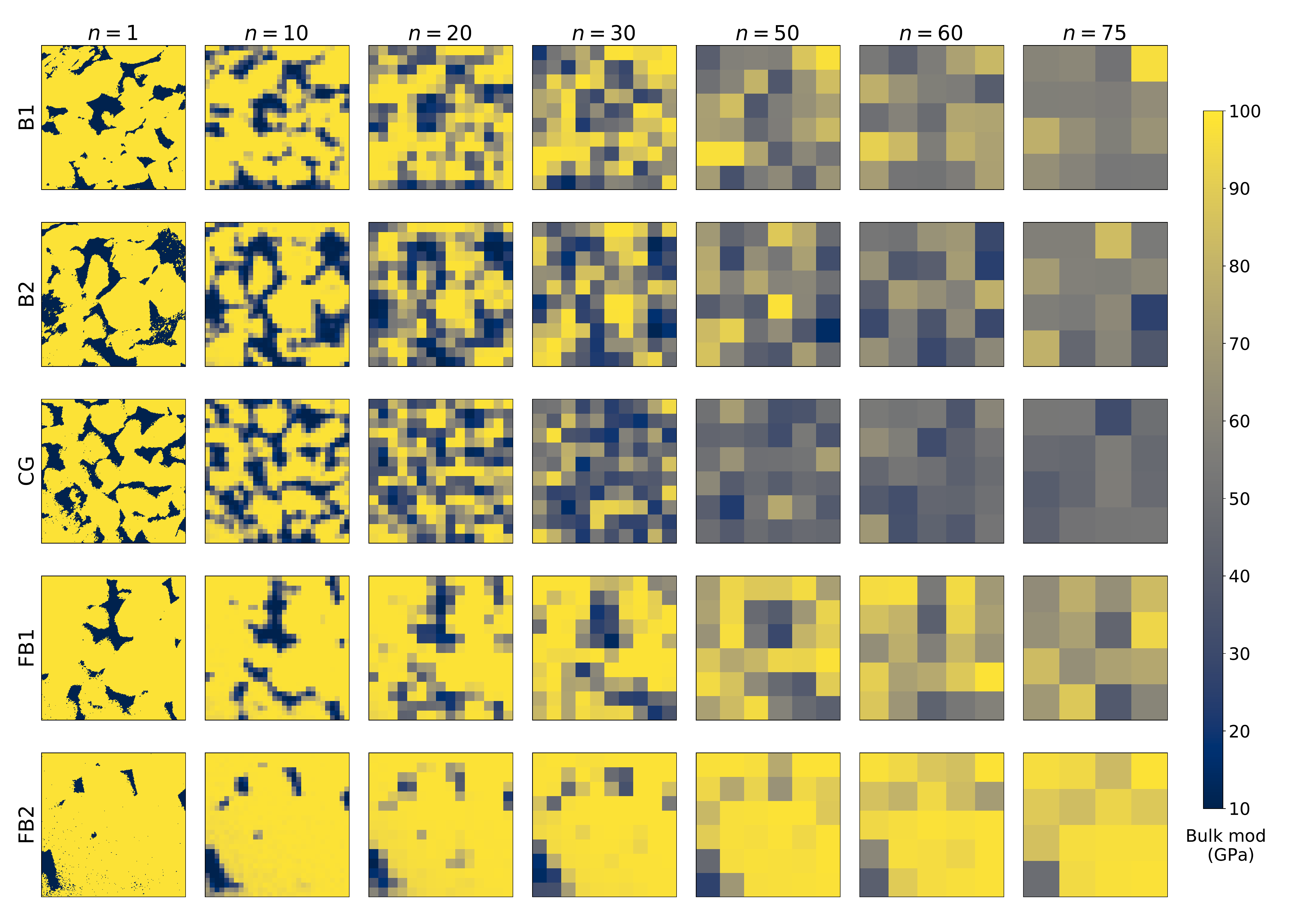}
    \caption{Distribution of bulk modulus in the assembled rock images after first step of hierarchical homogenization using different sizes $n$ of smaller images. A random two-dimensional slice of the three-dimensional image is plotted for each case. The rows correspond to different rock samples and the column are for different sizes $n$ of the small image used in the hierarchical homogenization. The first column ($n$ = 1) presents the bulk modulus distribution in the original image without any homogenization. We observe that increasing the size of the smaller image increases the homogeneity of the assembled image.}
    \label{fig_homo_size}
\end{figure}
The main finding of this work is the $n^{-1}$ scaling of the systematic error in the final homogenized elastic constants with the dimension $n$ of the  subimages. The error is caused mainly by two following reasons. The first is the complete neglect of the effect of surrounding materials on the homogenized elastic constants of the smaller subimages. The second is the mismatch between the boundary conditions used for the subimages during the first homogenization step and in the original image subject to its own loading. The homogenization of the subimages is performed under periodic boundary condition, however, the subimages are generally under mixed boundary conditions when they are part of the original image. Thus, the error arises from the inaccurate description of materials near the surface of the smaller subimages. This rationalizes the observed exponent of $-1$ associated with the power-law scaling of the error in hierarchical homogenization. This power law scaling, furthermore, enables a good estimate of the average elastic constants of the original big image. All these properties hold even for a different set of elastic properties of the two phases of the microstructure as shown in \ref{app:hh_result_2}.

The understanding of the cause of the systematic error in hierarchical homogenization also suggests a way to reduce this error. To incorporate the effect of the surrounding material and mimic the boundary condition experienced by the subimage as part of the big image, we introduce a padding of surrounding materials around the subimages during its homogenization. The error in the hierarchically homogenized elastic moduli progressively reduces with increase in the thickness of padding materials. The power law scaling of the hierarchically homogenized elastic moduli, with the exponent of $-1$ and almost same asymptotic value, is preserved across all the values of the padding thickness. The convergence of the hierarchically homogenized elastic moduli to the asymptotic value with respect to the subimage size accelerates with increase in the padding thickness. This further bolsters the robustness of the use of the power-law to extract the unbiased estimate of the elastic moduli of a big image using the hierarchical homogenization method.

The computational homogenization of the small subimages are performed by solving the local elasticity problem numerically using FFT-based elasticity solver. While the FFT-based elasticity solvers are efficient and fast, they satisfy equilibrium and compatibility conditions only for odd-sized grids, i.e., images with odd number of voxels. If the number of grid points is even, both conditions can not be satisfied simultaneously due to the treatment of the Nyquist frequency component ~\citep{Vondrejc2015, DeGeus2017a}. This difference in the odd- and even-sized (particularly, size of the intermediate assembled images, i.e. $N/n$) is reflected in the slightly jagged pattern of the blue line in Figures~\ref{fig_result_900} and~\ref{fig_result_900_pad}.

The size of the subimages to partition the original big image is chosen so that each subimage is sufficiently larger than the correlation length of the underlying microstructure. Thus, the elastic constants of the voxels of the assembled image are distributed about a mean value. The variance of the elastic constants decreases with increase in the subimage size rendering assembled image increasingly homogeneous, as shown in Figure~\ref{fig_homo_size}. This fact is also the basis of the trend of increasing error with decreasing subimage size. However, this trend does not hold as we continue to decrease the size of subimage. As illustrated in Figure~\ref{fig_homo_size}, once the subimage size is comparable to the correlation of the microstructure, the assembled image begins to retain the microstructural details and resemble the heterogeneous original image. Thus, below a cutoff size, the trend in error vs subimage size reverses, resulting in decrease in error with decrease in subimage size.

In conclusion, this work shows that the hierarchical homogenization method in combination with the power law scaling of the associated error can be employed to homogenize a huge domain of random heterogeneous materials such as rocks. This will enable the efficient computation of effective properties of very high-resolution images of the underlying heterogeneous materials.

\section*{Acknowledgement}
We acknowledge Shell for financial support and for providing the digital rock images. We thank Dr. Nishank Saxena at Shell for his constant support of the project. The authors also would like to thank Math2Market for providing the GeoDict software at a discount, and providing technical support.

\appendix
\section{FFT-based solver for elasticity problem}
\label{app_fft_solver}

The most important component of the computational homogenization is solving the equilibrium and compatibility equations~\eqref{eq_equilibrium_compatibility}. Since micro-CT scan and subsequent segmentation provides information about the rock microstructure in a three-dimensional grid, use of FFT-based solver is well-suited for solving the local elasticity problem with periodic boundary condition. The FFT-based solver was introduced by~\citet{Moulinec1994} and further expanded and advanced by several works~\citep{Moulinec1998a, Kabel2014, Vondrejc2014a, Vondrejc2015, DeGeus2017a,Zeman2017, Leute2022}. This section describes the basic formulation of the FFT-based solver, borrowing from Ref.~\citep{Zeman2017} which closely follows the standard finite element method (FEM) approach to obtain the discretized form of Equations~\eqref{eq_equilibrium_compatibility}. For a more comprehensive review of FFT-based homogenization, readers are referred to Refs~\citep{Schneider2021, Lucarini2022}

The discretization of local problem begins with reformulating the local problem, Equation~\eqref{eq_equilibrium_compatibility}, into the weak form, i.e.
\begin{align}
    \int_{\Omega} d\bs{x} ~\delta \varepsilon^*_{ij}(\bs{x}) ~\sigma_{ij}(\bs{x}, E_{ij} + \varepsilon^*_{ij}(\bs{x})) = 0,
\end{align}
where $\delta\varepsilon^*_{ij}(\bs{x})$ is a test strain field which itself must satisfy the compatibility and periodic boundary conditions.

The FEM approach satisfies the compatibility requirement and periodic boundary condition of the test strain field by expressing it in terms of a periodic test displacement field, $\delta \varepsilon^*_{ij}(\bs{x}) = (\delta u^*_{i,j}(\bs{x}) + \delta u^*_{j,i}(\bs{x}))/2$. The FFT-based approach, on the other hand, works directly with strain field without expressing it in terms of the underlying displacement field. In this case the compatibility of the test strain field is enforced via a four-rank self adjoint projection operator $\mathbb{G}_{ijkl}(\bs{x})$ which project any symmetric two-rank non-compatible tensor field $\zeta_{ij} (\bs{x})$ to a compatible part, i.e.
\begin{align}
    \delta \varepsilon_{ij}^*(\bs{x}) = \left[\mathbb{G} \star \bs{\zeta}\right]_{ij} (\bs{x}) = \int_{\Omega} d\bs{y} ~\mathbb{G}_{ijlk}(\bs{x}-\bs{y}) ~\zeta_{kl} (\bs{y})
\end{align}
is a compatible test strain field obtained by projecting non-compatible symmetric test field $\zeta_{ij}(\bs{x})$. In the above expression, $\star$ denotes convolution operation. The projection operator has a block diagonal form in the Fourier space, and is given by
\begin{equation}
    \begin{aligned}
        \hat{\mathbb{G}}_{ijlm}(\bs{k}) = \delta_{im} \hat{g}_{jl}(\bs{k}), \text{ where} \\
        \hat{g}_{jl}(\bs{k}) = \begin{cases}
            0                                          & \text{ if } \bs{k} = \bs{0} \\
            \frac{k_j k_l}{ \left\| \bs{k} \right\|^2} & \text{otherwise},
        \end{cases}
    \end{aligned}
\end{equation}
where the hat symbol $\hat{.}$ over a quantity denotes the fourier transform of that quantity; $\bs{k}$ is the frequency. The convolution operation is then performed efficiently in the Fourier space as
\begin{align}
    \left[\mathbb{G}\star \bs{\zeta}\right]_{ij}(\bs{x}) = \mathcal{F}^{-1}\left[\hat{\mathbb{G}}_{ijkl}(\bs{k}) ~\hat{\zeta}_{lk}(\bs{k}) \right](\bs{x}),
\end{align}
where $\mathcal{F}^{-1}[\cdot]$ is inverse Fourier transform operation.

Thus, the weak form can be written as
\begin{align}
    \int_{\Omega} d\bs{x} ~\left[\mathbb{G} \star \bs{\zeta} \right]_{ij} (\bs{x}) ~\sigma_{ij}(\bs{x}, E_{ij} + \varepsilon^*_{ij}(\bs{x})) = \int_{\Omega} d\bs{x} ~\zeta_{ij}(\bs{x}) ~\left[\mathbb{G}\star \bs{\sigma}\right]_{ij} (\bs{x}, E_{ij} + \varepsilon^*_{ij}(\bs{x})) = 0,
    \label{eq_weak_form_proj}
\end{align}
where the self-adjoint property of projection operator $\mathbb{G}_{ijkl}(\bs{x})$ is exploited to obtain the second expression of Equation~\eqref{eq_weak_form_proj} which has the advantage that the test field $\zeta_{ij}(\bs{x})$ now does not need to satisfy the compatibility condition and can be a member of the set of symmetric and periodic rank-two tensor fields.

\begin{figure}[ht!  ]
    \centering
    \includegraphics[width=0.3\textwidth]{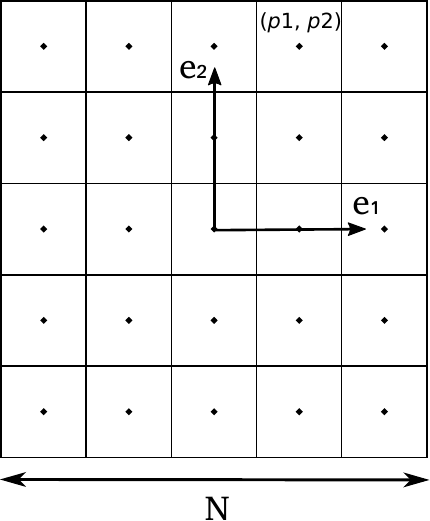}
    \caption{A schematic illustration of the discretization of the image and the coordinate system used in the FFT elasticity solver in two dimensions. The coordinate system is placed at the center of the image discretized in $N$ number of pixels in each direction. The center of each pixel is the discretization point. A discretization point is characterized by two integers $p_1$ and $p_2$, where $-N/2 < p_1, p_2 < N/2$.}
    \label{fig_ff_discretization}
\end{figure}

Equation~\eqref{eq_weak_form_proj} is now poised for discretization in the vein of the standard Galerkin method. To illustrate the discretization procedure, we focus on a two-dimensional problem. As shown in Figure~\ref{fig_ff_discretization}, we consider that the information about the microstructure is provided in the form of a square grid of size $N\times N$, i.e. $N$ nodes along each coordinate axis. Microstructure data obtained from micro-CT scan has this underlying grid structure (three-dimensional), and therefore belongs to the category under consideration. We further suppose that the number of pixels along each dimension $N$ is odd; this will simplify the exposition of the procedure by focusing on the essential features of FFT-based methods. The case of even number of pixels $N$ will be discussed later.

The coordinate of a grid point is  identified by a pair of numbers $\bs{p} = [p_1, p_2]$ as
\begin{align}
    \bs{x}^{\bs{p}}_N = p_1 \bs{e}_1 + p_2 \bs{e}_2,
\end{align}
where index $\bs{p}$ is sampled from a reduced set of index $\mathbb{Z}^2_N$, which is a set of pair of integers, each constrained to lie between $-N/2$ and $N/2$, i.e.
\begin{align}
    \bs{p} \in \mathbb{Z}^2_{N} = \left\{ \mathbb{Z}^2\left| -\frac{N}{2} \le p_1 \le \frac{N}{2}, -\frac{N}{2} \le p_1 \le \frac{N}{2} \right. \right\}.
\end{align}
The next step in the discretization of the weak form, Equation~\eqref{eq_weak_form_proj}, is definition of basis functions. In FFT-based methods, fundamental trigonometric polynomials are used as the basis functions. The fundamental trigonometric polynomial associated with node $\bs{p}$ is defined as
\begin{align}
    \varphi^{\bs{p}}_N(\bs{x}) = \frac{1}{N^2} \sum_{\bs{q}\in \mathbb{Z}^2_N} \exp\left( -\frac{2\pi i}{N} \bs{p}\cdot \bs{q} \right) \exp\left( \frac{2\pi i}{N} \bs{q}\cdot \bs{x} \right) \qquad \forall \bs{q} \in \mathbb{Z}^2_N.
\end{align}
The fundamental trigonometric polynomials, unlike FEM basis function, have global support, and still possess suitable interpolation properties, such as
\begin{equation}
    \begin{aligned}
        \varphi^{\bs{p}}_N \left(\bs{x}^{\bs{q}}_N\right)                        & = \delta_{\bs{km}},                   \\
        \sum_{\bs{p} \in \mathbb{Z}^2_N} \varphi^{\bs{p}}_N\left( \bs{x} \right) & = 1 \qquad \forall \bs{x} \in \Omega.
    \end{aligned}
    \label{eq_trig_poly_props}
\end{equation}
The first identity in above equation states that the fundamental trigonometric polynomial takes value 1 at the associated node, and it assumes value 0 at every other node. The second identity is a statement of the partition of unity.

Discretization of Equation~\eqref{eq_weak_form_proj} is performed by expressing both test field and the solution fluctuating strain field as linear combinations of the fundamental trigonometric polynomials
\begin{equation}
    \begin{aligned}
        \zeta_{ij}(\bs{x})         & = \sum_{\bs{p}\in \mathbb{Z}^2_N} \varphi^{\bs{p}}_N(\bs{x})~\zeta_{ij}^{\bs{p}},        \\
        \varepsilon^*_{ij}(\bs{x}) & = \sum_{\bs{q}\in \mathbb{Z}^2_N} \varphi^{\bs{q}}_N(\bs{x})~\varepsilon^{*\bs{q}}_{ij},
    \end{aligned}
    \label{eq_test_strain_expansion}
\end{equation}
where $\zeta_{ij}^{\bs{p}}$ and $\varepsilon^{*\bs{p}}_{ij}$ are values of the test and strain field, respectively, at node $\bs{p}$. We then insert Equation~\eqref{eq_test_strain_expansion} into Equation~\eqref{eq_weak_form_proj} to obtain
\begin{equation}
    \begin{aligned}
        \int_{\Omega} d\bs{x} \sum_{\bs{p}\in \mathbb{Z}^2_N} \varphi^{\bs{p}}_N(\bs{x})~\zeta_{ij}^{\bs{p}} ~\left[\mathbb{G}\star \bs{\sigma}\right]_{ij} \left(\bs{x}, E_{ij} + \sum_{\bs{q}\in \mathbb{Z}^2_N} \varphi^{\bs{q}}_N(\bs{x})~\varepsilon^{*\bs{q}}_{ij}\right) \approx 0,                                         \\
        \sum_{\bs{m}\in\mathbb{Z}^2} \sum_{\bs{p}\in \mathbb{Z}^2_N} \varphi^{\bs{p}}_N(\bs{x}^{\bs{m}}_N)~\zeta_{ij}^{\bs{p}} ~\left[\mathbb{G}\star \bs{\sigma}\right]_{ij} \left(\bs{x}^{\bs{m}}_N, E_{ij} + \sum_{\bs{q}\in \mathbb{Z}^2_N} \varphi^{\bs{q}}_N(\bs{x}^{\bs{m}}_N)~\varepsilon^{*\bs{q}}_{ij}\right) \approx 0, \\
        \sum_{\bs{m}\in\mathbb{Z}^2} \zeta_{ij}^{\bs{m}} ~\left[\mathbb{G}\star \bs{\sigma}\right]_{ij} \left(\bs{x}^{\bs{m}}_N, E_{ij} +\varepsilon^{*\bs{m}}_{ij}\right) \approx 0.
    \end{aligned}
    \label{eq_discrete_man_weak}
\end{equation}
The second line of the equation is obtained by numerically integrating the first line using the trapezoidal quadrature rule. The third line is the result of the application of the first property of the fundamental trigonometric polynomials as presented in Equation~\eqref{eq_trig_poly_props}. The final step in the discretization is application of the fact that the third line of Equation~\eqref{eq_discrete_man_weak} is satisfied for all admissible values of the test fields $\zeta_{ij}^{\bs{m}}$. This condition leads to the following set of equations
\begin{align}
    \left[\mathbb{G}\star \bs{\sigma}\right]_{ij} \left(\bs{x}^{\bs{m}}_N, E_{ij} +\varepsilon^{*\bs{m}}_{ij}\right) \approx 0 \qquad \forall \bs{m} \in \mathbb{Z}^2_N.
\end{align}
This equation is further simplified by the use of local constitutive relation for each node. Since material associated with each node is assumed to be linear isotropic elastic with its own elastic constant tensor, we reach to following equation
\begin{align}
    \mathbb{G} \star \left(\mathbb{C}^{\bs{m}} : \bs{\varepsilon}^{*\bs{m}} \right)   = - \mathbb{G} \star \left(\mathbb{C}^{\bs{m}} : \bs{E} \right),
\end{align}
where $\mathbb{C}^*$ is elastic constant tensor of the material associated with node $m$. This is a linear equation in nodal fluctuation strain $\varepsilon_{ij}^{*m}$, and is solved by using conjugate gradient method. The use of conjugate gradient method also ensures compatibility of $\varepsilon_{ij}^{*\bs{m}}$ field at every step of the iteration. For further implementation details on FFT-based elasticity solver, interested readers are referred to Refs~\citep{DeGeus2017a,Zeman2017}.

The main advantages of the FFT-based elasticity solvers include: (1) The solver operates directly on the images of microstructure and obviate the need for generating meshes to represent the microstructure, (2)  the global stiffness matrix need not be assembled, (3) use of efficient fast fourier transform algorithm to perform the convolution operation, and (4) natural integration of periodic boundary conditions which is desirable for homogenization problems.

\section{Microstructure with elastic properties:$K_1 = 98.04$~GPa, $\mu_1 = 37.59$~GPa; $K_2 = 9.8.0$~GPa, $\mu_2 = 3.76$~GPa }
\label{app:hh_result_2}
In this section, we present the results of hierarchical homogenization of the five microstructures where phase 1 and phase 2 are assigned elastic properties different than those presented in the main text. In particular, bulk and shear moduli of phase 1 are, respectively, 98.04~GPa and 37.59~GPa; and bulk and shear moduli of phase 1 are, respectively, 9.8.0~GPa and 3.76~GPa. Thus, phase 1 is ten times stiffer than phase 2.

\begin{figure}[H]
    \centering
    \includegraphics[width=\textwidth]{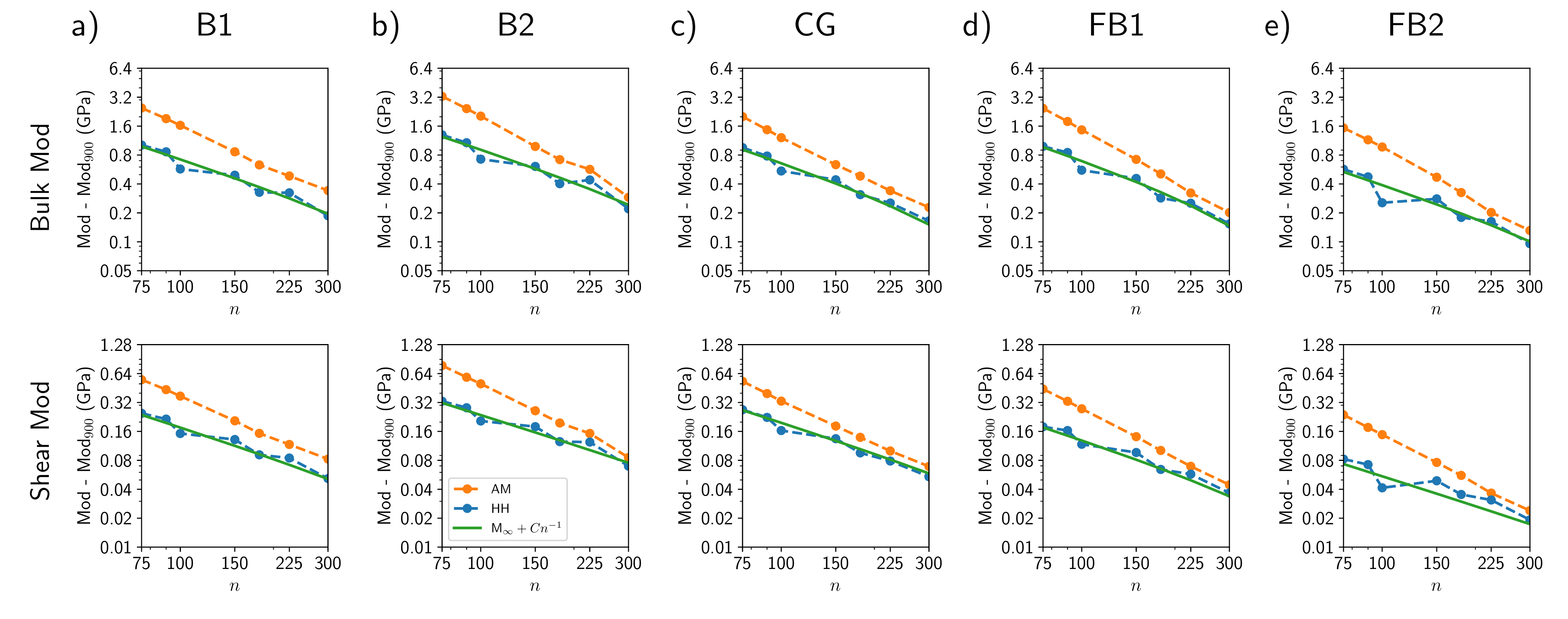}
    \caption{Effect of subimage size on the hierarchically homogenized elastic constants for zero padding case for the five rock samples of size $N = 900$: (a) B1, (b) B2, (c) CG, (d) FB1, and (e) FB2. In each subfigure, corresponding to a specific rock sample, the variation of errors in the homogenized elastic moduli obtained from the hierarchical homogenization are plotted on log-log scale as a function of the subimage size $n$. The error is calculated by subtracting the corresponding homogenized elastic moduli of the big image of size $n=900$ that is obtained by solving the local elasticity problem in the big image. The first panel in each subfigure shows the result for the bulk modulus and the second panel for the shear modulus. The orange plots show the arithmetic mean (AM) of the elastic moduli of the small images composing the big image; the blue plots show the result obtained from the hierarchical homogenization (HH); and the green line is the fit of the hierarchical homogenization values to the function $M_{\infty} + C n^{-1}$, with $M_\infty$ and $C$ being the two fitting parameters.}
    \label{fig_result_900_2}
\end{figure}

\begin{figure}[H]
    \centering
    \includegraphics[width=\textwidth]{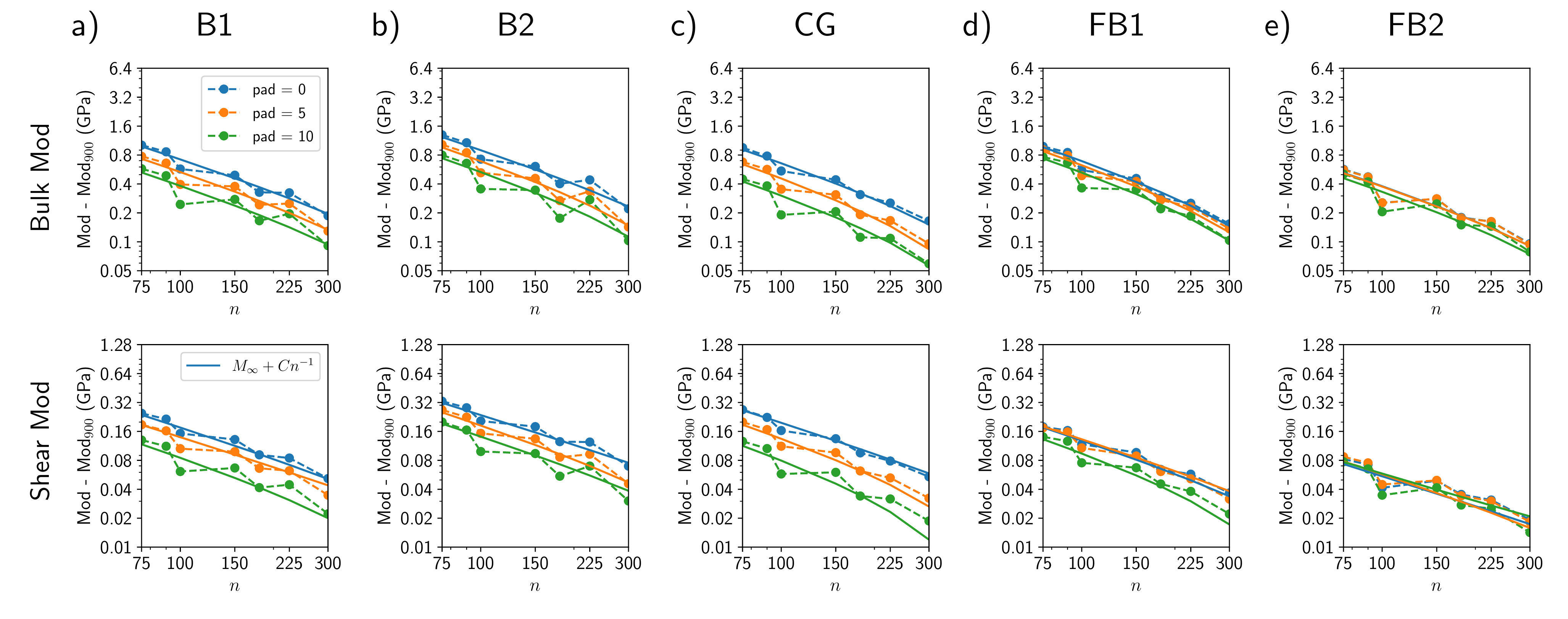}
    \caption{Effect of the padding thickness on the hierarchically homogenized elastic constants for different padding thickness for the five rock samples of size $N = 900$: (a) B1, (b) B2, (c) CG, (d) FB1, and (e) FB2. In each subfigure, corresponding to a specific rock sample, the variation of errors in the homogenized elastic moduli obtained from the hierarchical homogenization are plotted on log-log scale as a function of the subimage size $n$ for the three values of padding thickness: 0, 10, 20. The error is calculated by subtracting the corresponding homogenized elastic moduli of the big image of size $n=900$ that is obtained by solving the local elasticity problem in the big image. The first panel in each subfigure shows the result for the bulk modulus and the second panel for the shear modulus. The three colors correspond to the three values of the padding thickness: blue for 0, orange for 5, and green for 10. The dashed lines with circle markers show the simulation result and the solid curve is the fit of the hierarchical homogenization values to the function $M_{\infty} + C n^{-1}$, with $M_\infty$ and $C$ being the two fitting parameters. Across all the five rocks, the error decreases with increasing padding thickness for a fixed subimage size.}
    \label{fig_result_900_pad_2}
\end{figure}

Figure~\ref{fig_result_900_2} presents the homogenized elastic moduli obtained from the hierarchical homogenization and arithmetic average for zero padding case. Figure~\ref{fig_result_900_pad_2} shows the effects of padding thickness of the surrounding materials around the subimages on the hierarchically homogenized elastic moduli. Furthermore, the values obtained from these computations are presented in Table~\ref{table_result_hh_900_2} and~\ref{table_fit_results_900_2}. The overall behaviors of these results are the same to that presented in the main text where the two phases have the properties of quartz and pore. In particular, we observe that hierarchically homogenized elastic moduli follow the power law with the exponent $-1$ across all padding thickness  values (0, 5, 10) studied; error in the hierarchically homogenized elastic moduli reduces with padding thickness for a fixed subimage size $n$; convergence of hierarchically homogenized elastic moduli to the corresponding asymptotic value with subimage size $n$ becomes faster with increasing padding thickness. Thus, the emergence of the power law and effect of padding materials are expected to be robust and depends upon the particular values of the phase properties.

\begin{table}[H]
    \centering
    \caption{ Homogenized elastic moduli (bulk and shear moduli) of the five rock images of size $N=900$ as a function of small image size $n$ into which the big image is partitioned. The values are obtained by the hierarchical homogenization method and by taking arithmetic mean of the homogenized elastic moduli of the small images. }
    \begin{tabular}{l c  c c c c c c c c  }
        \toprule
        \multicolumn{3}{c}{\diagbox{Rock type}{$n$}} & 75                               & 90 & 100   & 150   & 180   & 225   & 300                   \\
        \midrule
        \multirow{4}{*}{B1}                          & \multirow{2}{*}{Bulk mod (GPa)}  & HH & 61.48 & 61.33 & 61.04 & 60.96 & 60.79 & 60.79 & 60.65 \\
                                                     &                                  & AM & 62.93 & 62.38 & 62.10 & 61.33 & 61.10 & 60.95 & 60.81 \\
        \cline{ 3-10 }
                                                     & \multirow{2}{*}{Shear mod (GPa)} & HH & 26.92 & 26.89 & 26.83 & 26.81 & 26.77 & 26.76 & 26.73 \\
                                                     &                                  & AM & 27.23 & 27.11 & 27.05 & 26.88 & 26.83 & 26.79 & 26.76 \\
        \midrule
        \multirow{4}{*}{B2}                          & \multirow{2}{*}{Bulk mod (GPa)}  & HH & 56.61 & 55.31 & 56.04 & 55.92 & 55.71 & 55.75 & 55.53 \\
                                                     &                                  & AM & 58.58 & 57.74 & 57.34 & 56.29 & 56.03 & 55.88 & 55.60 \\
        \cline{ 3-10 }
                                                     & \multirow{2}{*}{Shear mod (GPa)} & HH & 25.20 & 24.87 & 25.07 & 25.05 & 24.99 & 24.99 & 24.93 \\
                                                     &                                  & AM & 25.64 & 25.45 & 25.37 & 25.13 & 25.06 & 25.02 & 24.95 \\
        \midrule
        \multirow{4}{*}{CG}                          & \multirow{2}{*}{Bulk mod (GPa)}  & HH & 52.82 & 52.65 & 52.41 & 52.31 & 52.18 & 52.12 & 52.03 \\
                                                     &                                  & AM & 53.88 & 53.34 & 53.08 & 52.50 & 52.35 & 52.21 & 52.10 \\
        \cline{ 3-10 }
                                                     & \multirow{2}{*}{Shear mod (GPa)} & HH & 23.88 & 23.83 & 23.77 & 23.75 & 23.71 & 23.69 & 23.67 \\
                                                     &                                  & AM & 24.14 & 24.00 & 23.94 & 23.79 & 23.75 & 23.71 & 23.68 \\
        \midrule
        \multirow{4}{*}{FB1}                         & \multirow{2}{*}{Bulk mod (GPa)}  & HH & 75.62 & 75.49 & 75.19 & 75.09 & 74.92 & 74.89 & 74.79 \\
                                                     &                                  & AM & 77.09 & 76.42 & 76.10 & 75.36 & 75.15 & 74.96 & 74.84 \\
        \cline{ 3-10 }
                                                     & \multirow{2}{*}{Shear mod (GPa)} & HH & 31.42 & 31.41 & 31.36 & 31.34 & 31.31 & 31.30 & 31.28 \\
                                                     &                                  & AM & 31.69 & 31.57 & 31.52 & 31.38 & 31.34 & 31.31 & 31.29 \\
        \midrule
        \multirow{4}{*}{FB2}                         & \multirow{2}{*}{Bulk mod (GPa)}  & HH & 89.12 & 89.02 & 88.80 & 88.82 & 88.73 & 88.71 & 88.64 \\
                                                     &                                  & AM & 90.08 & 89.70 & 89.52 & 89.02 & 88.87 & 88.75 & 88.68 \\
        \cline{ 3-10 }
                                                     & \multirow{2}{*}{Shear mod (GPa)} & HH & 35.25 & 35.24 & 35.21 & 35.21 & 35.20 & 35.19 & 35.18 \\
                                                     &                                  & AM & 35.40 & 35.34 & 35.31 & 35.24 & 35.22 & 35.20 & 35.19 \\
        \bottomrule
    \end{tabular}
    \label{table_result_hh_900_2}
\end{table}

\begin{table}[H]
    \centering
    \caption{Homogenized properties of the five rock samples. $M_{900}$ denotes the elastic modulus obtained by directly solving the big image of size $N=900$. $M_{\infty}$ is the asymptotic value of the corresponding elastic modulus that is obtained by fitting the hierarchical homogenization result to the power law given in Equation~\eqref{eq_fit_power_law}. The value of $M_\infty$ is a good approximation for the homogenized elastic modulus of a big rock sample.}
    \begin{tabular}{l  c   c  c  c   c  c c  }
        \toprule
        \multirow{2}{*}{Rock type} & \multirow{2}{*}{Porosity(\%)} & \multicolumn{3}{c}{Bulk modulus (GPa)} & \multicolumn{3}{ c }{Shear modulus (GPa)}                                            \\
        \cline{3-8}
                                   &                               & $M_{900}$                              & $M_{\infty}$                              & $C$   & $M_{900}$ & $M_{\infty}$ & $C$   \\
        \midrule
        B1                         & 16.51                         & 60.47                                  & 60.40                                     & 78.87 & 26.67     & 26.68        & 18.64 \\
        B2                         & 19.63                         & 55.21                                  & 55.31                                     & 99.77 & 24.86     & 24.87        & 24.18 \\
        CG                         & 22.20                         & 51.77                                  & 51.87                                     & 75.83 & 23.60     & 23.61        & 20.73 \\
        FB1                        & 9.25                          & 74.51                                  & 74.64                                     & 81.89 & 31.23     & 31.24        & 14.22 \\
        FB2                        & 3.45                          & 88.50                                  & 88.55                                     & 43.24 & 35.15     & 35.16        & 55.84 \\
        \bottomrule
    \end{tabular}
    \label{table_fit_results_900_2}
\end{table}

\addcontentsline{toc}{section}{References}
\bibliographystyle{unsrt}

\begin{thebibliography}{51}
\expandafter\ifx\csname natexlab\endcsname\relax\def\natexlab#1{#1}\fi
\providecommand{\bibinfo}[2]{#2}
\ifx\xfnm\relax \def\xfnm[#1]{\unskip,\space#1}\fi
\bibitem[{Fredrich et~al.(1993)Fredrich, Greaves, and Martin}]{Fredrich1993}
\bibinfo{author}{J.~T. Fredrich}, \bibinfo{author}{K.~H. Greaves},
  \bibinfo{author}{J.~W. Martin},
\newblock \bibinfo{title}{{Pore geometry and transport properties of
  Fontainebleau sandstone}},
\newblock \bibinfo{journal}{Int. J. Rock Mech. Min. Sci.} \bibinfo{volume}{30}
  (\bibinfo{year}{1993}) \bibinfo{pages}{691--697}.
\bibitem[{Arns et~al.(2001)Arns, Knackstedt, Pinczewski, and {W. B.
  Lindquist}}]{Arns2001}
\bibinfo{author}{C.~H. Arns}, \bibinfo{author}{M.~A. Knackstedt},
  \bibinfo{author}{W.~V. Pinczewski}, \bibinfo{author}{{W. B. Lindquist}},
\newblock \bibinfo{title}{{Accurate estimation of transport properties from
  microtomographic images}},
\newblock \bibinfo{journal}{Geophys. Res. Lett.} \bibinfo{volume}{28}
  (\bibinfo{year}{2001}) \bibinfo{pages}{3361--3364}.
\bibitem[{Dvorkin et~al.(2011)Dvorkin, Derzhi, Diaz, and Fang}]{Dvorkin2011}
\bibinfo{author}{J.~Dvorkin}, \bibinfo{author}{N.~Derzhi},
  \bibinfo{author}{E.~Diaz}, \bibinfo{author}{Q.~Fang},
\newblock \bibinfo{title}{{Relevance of computational rock physics}},
\newblock \bibinfo{journal}{Geophysics} \bibinfo{volume}{76}
  (\bibinfo{year}{2011}).
\bibitem[{Andr{\"{a}} et~al.(2013{\natexlab{a}})Andr{\"{a}}, Combaret, Dvorkin,
  Glatt, Han, Kabel, Keehm, Krzikalla, Lee, Madonna, Marsh, Mukerji, Saenger,
  Sain, Saxena, Ricker, Wiegmann, and Zhan}]{Andra2013}
\bibinfo{author}{H.~Andr{\"{a}}}, \bibinfo{author}{N.~Combaret},
  \bibinfo{author}{J.~Dvorkin}, \bibinfo{author}{E.~Glatt},
  \bibinfo{author}{J.~Han}, \bibinfo{author}{M.~Kabel},
  \bibinfo{author}{Y.~Keehm}, \bibinfo{author}{F.~Krzikalla},
  \bibinfo{author}{M.~Lee}, \bibinfo{author}{C.~Madonna},
  \bibinfo{author}{M.~Marsh}, \bibinfo{author}{T.~Mukerji},
  \bibinfo{author}{E.~H. Saenger}, \bibinfo{author}{R.~Sain},
  \bibinfo{author}{N.~Saxena}, \bibinfo{author}{S.~Ricker},
  \bibinfo{author}{A.~Wiegmann}, \bibinfo{author}{X.~Zhan},
\newblock \bibinfo{title}{{Digital rock physics benchmarks-Part I: Imaging and
  segmentation}},
\newblock \bibinfo{journal}{Comput. Geosci.} \bibinfo{volume}{50}
  (\bibinfo{year}{2013}{\natexlab{a}}) \bibinfo{pages}{25--32}.
\bibitem[{Andr{\"{a}} et~al.(2013{\natexlab{b}})Andr{\"{a}}, Combaret, Dvorkin,
  Glatt, Han, Kabel, Keehm, Krzikalla, Lee, Madonna, Marsh, Mukerji, Saenger,
  Sain, Saxena, Ricker, Wiegmann, and Zhan}]{Andra2013a}
\bibinfo{author}{H.~Andr{\"{a}}}, \bibinfo{author}{N.~Combaret},
  \bibinfo{author}{J.~Dvorkin}, \bibinfo{author}{E.~Glatt},
  \bibinfo{author}{J.~Han}, \bibinfo{author}{M.~Kabel},
  \bibinfo{author}{Y.~Keehm}, \bibinfo{author}{F.~Krzikalla},
  \bibinfo{author}{M.~Lee}, \bibinfo{author}{C.~Madonna},
  \bibinfo{author}{M.~Marsh}, \bibinfo{author}{T.~Mukerji},
  \bibinfo{author}{E.~H. Saenger}, \bibinfo{author}{R.~Sain},
  \bibinfo{author}{N.~Saxena}, \bibinfo{author}{S.~Ricker},
  \bibinfo{author}{A.~Wiegmann}, \bibinfo{author}{X.~Zhan},
\newblock \bibinfo{title}{{Digital rock physics benchmarks-part II: Computing
  effective properties}},
\newblock \bibinfo{journal}{Comput. Geosci.} \bibinfo{volume}{50}
  (\bibinfo{year}{2013}{\natexlab{b}}) \bibinfo{pages}{33--43}.
\bibitem[{Saxena et~al.(2017)Saxena, Hofmann, Alpak, Dietderich, Hunter, and
  Day-Stirrat}]{Saxena2017}
\bibinfo{author}{N.~Saxena}, \bibinfo{author}{R.~Hofmann},
  \bibinfo{author}{F.~O. Alpak}, \bibinfo{author}{J.~Dietderich},
  \bibinfo{author}{S.~Hunter}, \bibinfo{author}{R.~J. Day-Stirrat},
\newblock \bibinfo{title}{{Effect of image segmentation {\&} voxel size on
  micro-CT computed effective transport {\&} elastic properties}},
\newblock \bibinfo{journal}{Mar. Pet. Geol.} \bibinfo{volume}{86}
  (\bibinfo{year}{2017}) \bibinfo{pages}{972--990}.
\bibitem[{Saxena et~al.(2019)Saxena, Hofmann, Hows, Saenger, Duranti, Stefani,
  Wiegmann, Kerimov, and Kabel}]{Saxena2019}
\bibinfo{author}{N.~Saxena}, \bibinfo{author}{R.~Hofmann},
  \bibinfo{author}{A.~Hows}, \bibinfo{author}{E.~H. Saenger},
  \bibinfo{author}{L.~Duranti}, \bibinfo{author}{J.~Stefani},
  \bibinfo{author}{A.~Wiegmann}, \bibinfo{author}{A.~Kerimov},
  \bibinfo{author}{M.~Kabel},
\newblock \bibinfo{title}{{Rock compressibility from microcomputed tomography
  images: Controls on digital rock simulations}},
\newblock \bibinfo{journal}{Geophysics} \bibinfo{volume}{84}
  (\bibinfo{year}{2019}) \bibinfo{pages}{WA127--WA139}.
\bibitem[{Sun et~al.(2021)Sun, Salazar-Tio, Duranti, Crouse, Fager, and
  Balasubramanian}]{Sun2021}
\bibinfo{author}{Z.~Sun}, \bibinfo{author}{R.~Salazar-Tio},
  \bibinfo{author}{L.~Duranti}, \bibinfo{author}{B.~Crouse},
  \bibinfo{author}{A.~Fager}, \bibinfo{author}{G.~Balasubramanian},
\newblock \bibinfo{title}{{Prediction of rock elastic moduli based on a
  micromechanical finite element model}},
\newblock \bibinfo{journal}{Comput. Geotech.} \bibinfo{volume}{135}
  (\bibinfo{year}{2021}) \bibinfo{pages}{104149}.
\bibitem[{Voigt(1910)}]{voigt1910lehrbuch}
\bibinfo{author}{W.~Voigt}, \bibinfo{title}{{Lehrbuch der kristallphysik: (mit
  ausschluss der kristalloptik)}}, B.G. Teubners Sammlung von Lehrb{\"{u}}chern
  auf dem Gebiete der mathematischen Wissenschaften ; Bd. XXXIV,
  \bibinfo{publisher}{B.G. Teubner}, \bibinfo{year}{1910}.
\bibitem[{Reuss(1929)}]{Reuss1929}
\bibinfo{author}{A.~Reuss},
\newblock \bibinfo{title}{{Berechnung der Flie{\ss}grenze von Mischkristallen
  auf Grund der Plastizit{\"{a}}tsbedingung f{\"{u}}r Einkristalle .}},
\newblock \bibinfo{journal}{ZAMM - J. Appl. Math. Mech. / Zeitschrift f{\"{u}}r
  Angew. Math. und Mech.} \bibinfo{volume}{9} (\bibinfo{year}{1929})
  \bibinfo{pages}{49--58}.
\bibitem[{Hashin and Shtrikman(1963)}]{Hashin1963}
\bibinfo{author}{Z.~Hashin}, \bibinfo{author}{S.~Shtrikman},
\newblock \bibinfo{title}{{A variational approach to the theory of the elastic
  behaviour of multiphase materials}},
\newblock \bibinfo{journal}{J. Mech. Phys. Solids} \bibinfo{volume}{11}
  (\bibinfo{year}{1963}) \bibinfo{pages}{127--140}.
\bibitem[{Nemat-Nasser et~al.(2013)Nemat-Nasser, Hori, and
  Achenbach}]{nemat2013micromechanics}
\bibinfo{author}{S.~Nemat-Nasser}, \bibinfo{author}{M.~Hori},
  \bibinfo{author}{J.~D. Achenbach}, \bibinfo{title}{{Micromechanics: Overall
  Properties of Heterogeneous Materials}}, ISSN, \bibinfo{publisher}{Elsevier
  Science}, \bibinfo{year}{2013}.
\bibitem[{Zaoui(2002)}]{Zaoui2002}
\bibinfo{author}{A.~Zaoui},
\newblock \bibinfo{title}{{Continuum Micromechanics: Survey}},
\newblock \bibinfo{journal}{J. Eng. Mech.} \bibinfo{volume}{128}
  (\bibinfo{year}{2002}) \bibinfo{pages}{808--816}.
\bibitem[{Charalambakis(2010)}]{Charalambakis2010}
\bibinfo{author}{N.~Charalambakis},
\newblock \bibinfo{title}{{Homogenization techniques and micromechanics. A
  survey and perspectives}},
\newblock \bibinfo{journal}{Appl. Mech. Rev.} \bibinfo{volume}{63}
  (\bibinfo{year}{2010}) \bibinfo{pages}{1--10}.
\bibitem[{Yvonnet(2019)}]{yvonnet2019computational}
\bibinfo{author}{J.~Yvonnet}, \bibinfo{title}{{Computational Homogenization of
  Heterogeneous Materials with Finite Elements}}, Solid Mechanics and Its
  Applications, \bibinfo{publisher}{Springer International Publishing},
  \bibinfo{year}{2019}.
\bibitem[{Khdir et~al.(2013)Khdir, Kanit, Za{\"{i}}ri, and
  Na{\"{i}}t-Abdelaziz}]{Khdir2013}
\bibinfo{author}{Y.~K. Khdir}, \bibinfo{author}{T.~Kanit},
  \bibinfo{author}{F.~Za{\"{i}}ri}, \bibinfo{author}{M.~Na{\"{i}}t-Abdelaziz},
\newblock \bibinfo{title}{{Computational homogenization of elastic-plastic
  composites}},
\newblock \bibinfo{journal}{Int. J. Solids Struct.} \bibinfo{volume}{50}
  (\bibinfo{year}{2013}) \bibinfo{pages}{2829--2835}.
\bibitem[{Takano et~al.(2000)Takano, Zako, and Ishizono}]{Takano2000}
\bibinfo{author}{N.~Takano}, \bibinfo{author}{M.~Zako},
  \bibinfo{author}{M.~Ishizono},
\newblock \bibinfo{title}{{Multi-scale computational method for elastic bodies
  with global and local heterogeneity}},
\newblock \bibinfo{journal}{J. Comput. Mater. Des.} \bibinfo{volume}{7}
  (\bibinfo{year}{2000}) \bibinfo{pages}{111--132}.
\bibitem[{Vel et~al.(2016)Vel, Cook, Johnson, and Gerbi}]{Vel2016}
\bibinfo{author}{S.~S. Vel}, \bibinfo{author}{A.~C. Cook},
  \bibinfo{author}{S.~E. Johnson}, \bibinfo{author}{C.~Gerbi},
\newblock \bibinfo{title}{{Computational homogenization and micromechanical
  analysis of textured polycrystalline materials}},
\newblock \bibinfo{journal}{Comput. Methods Appl. Mech. Eng.}
  \bibinfo{volume}{310} (\bibinfo{year}{2016}) \bibinfo{pages}{749--779}.
\bibitem[{Kadanoff(1966)}]{Kadanoff1966}
\bibinfo{author}{L.~P. Kadanoff},
\newblock \bibinfo{title}{{Scaling laws for Ising models near Tc}},
\newblock \bibinfo{journal}{Physics (College. Park. Md).} \bibinfo{volume}{2}
  (\bibinfo{year}{1966}) \bibinfo{pages}{263--272}.
\bibitem[{Wilson(1971)}]{Wilson1971}
\bibinfo{author}{K.~G. Wilson},
\newblock \bibinfo{title}{{Renormalization group and critical phenomena. I.
  Renormalization group and the Kadanoff scaling picture}},
\newblock \bibinfo{journal}{Phys. Rev. B} \bibinfo{volume}{4}
  (\bibinfo{year}{1971}) \bibinfo{pages}{3174--3183}.
\bibitem[{Wilson(1975)}]{Wilson1975}
\bibinfo{author}{K.~G. Wilson},
\newblock \bibinfo{title}{{The renormalization group: Critical phenomena and
  the Kondo problem}},
\newblock \bibinfo{journal}{Rev. Mod. Phys.} \bibinfo{volume}{47}
  (\bibinfo{year}{1975}) \bibinfo{pages}{773--840}.
\bibitem[{Wilson(1983)}]{Wilson1983}
\bibinfo{author}{K.~G. Wilson},
\newblock \bibinfo{title}{{The renormalization group and critical phenomena}},
\newblock \bibinfo{journal}{Rev. Mod. Phys.} \bibinfo{volume}{55}
  (\bibinfo{year}{1983}) \bibinfo{pages}{583--600}.
\bibitem[{Fisher(1998)}]{Fisher1998}
\bibinfo{author}{M.~E. Fisher},
\newblock \bibinfo{title}{{Renormalization group theory: Its basis and
  formulation in statistical physics}},
\newblock \bibinfo{journal}{Rev. Mod. Phys.} \bibinfo{volume}{70}
  (\bibinfo{year}{1998}) \bibinfo{pages}{653--681}.
\bibitem[{Efrati et~al.(2014)Efrati, Wang, Kolan, and Kadanoff}]{Efrati2014}
\bibinfo{author}{E.~Efrati}, \bibinfo{author}{Z.~Wang},
  \bibinfo{author}{A.~Kolan}, \bibinfo{author}{L.~P. Kadanoff},
\newblock \bibinfo{title}{{Real-space renormalization in statistical
  mechanics}},
\newblock \bibinfo{journal}{Rev. Mod. Phys.} \bibinfo{volume}{86}
  (\bibinfo{year}{2014}) \bibinfo{pages}{647--667}.
\bibitem[{Kim and Russel(1985)}]{Russel1985}
\bibinfo{author}{S.~Kim}, \bibinfo{author}{W.~B. Russel},
\newblock \bibinfo{title}{{Modelling of porous media by renormalization of the
  Stokes equations}},
\newblock \bibinfo{journal}{J. Fluid Mech.} \bibinfo{volume}{154}
  (\bibinfo{year}{1985}) \bibinfo{pages}{269--286}.
\bibitem[{King(1989)}]{King1989}
\bibinfo{author}{P.~R. King},
\newblock \bibinfo{title}{{The use of renormalization for calculating effective
  permeability}},
\newblock \bibinfo{journal}{Transp. Porous Media} \bibinfo{volume}{4}
  (\bibinfo{year}{1989}) \bibinfo{pages}{37--58}.
\bibitem[{King(1996)}]{King1996}
\bibinfo{author}{P.~R. King},
\newblock \bibinfo{title}{{Upscaling permeability: Error analysis for
  renormalization}},
\newblock \bibinfo{journal}{Transp. Porous Media} \bibinfo{volume}{23}
  (\bibinfo{year}{1996}) \bibinfo{pages}{337--354}.
\bibitem[{Banerjee and Adams(2004)}]{Banerjee2004}
\bibinfo{author}{B.~Banerjee}, \bibinfo{author}{D.~O. Adams},
\newblock \bibinfo{title}{{On predicting the effective elastic properties of
  polymer bonded explosives using the recursive cell method}},
\newblock \bibinfo{journal}{Int. J. Solids Struct.} \bibinfo{volume}{41}
  (\bibinfo{year}{2004}) \bibinfo{pages}{481--509}.
\bibitem[{Green and Paterson(2007)}]{Green2007}
\bibinfo{author}{C.~P. Green}, \bibinfo{author}{L.~Paterson},
\newblock \bibinfo{title}{{Analytical three-dimensional renormalization for
  calculating effective permeabilities}},
\newblock \bibinfo{journal}{Transp. Porous Media} \bibinfo{volume}{68}
  (\bibinfo{year}{2007}) \bibinfo{pages}{237--248}.
\bibitem[{Karim and Krabbenhoft(2010)}]{Karim2010}
\bibinfo{author}{M.~R. Karim}, \bibinfo{author}{K.~Krabbenhoft},
\newblock \bibinfo{title}{{New Renormalization Schemes for Conductivity
  Upscaling in Heterogeneous Media}},
\newblock \bibinfo{journal}{Transp. Porous Media} \bibinfo{volume}{85}
  (\bibinfo{year}{2010}) \bibinfo{pages}{677--690}.
\bibitem[{Hanasoge et~al.(2017)Hanasoge, Agarwal, Tandon, and
  Koelman}]{Hanasoge2017}
\bibinfo{author}{S.~Hanasoge}, \bibinfo{author}{U.~Agarwal},
  \bibinfo{author}{K.~Tandon}, \bibinfo{author}{J.~M.~A. Koelman},
\newblock \bibinfo{title}{{Renormalization group theory outperforms other
  approaches in statistical comparison between upscaling techniques for porous
  media}},
\newblock \bibinfo{journal}{Phys. Rev. E} \bibinfo{volume}{96}
  (\bibinfo{year}{2017}) \bibinfo{pages}{1--10}.
\bibitem[{Hansen et~al.(1997)Hansen, Roux, Aharony, Feder, J{\o}ssang, and
  Hardy}]{Hansen1997a}
\bibinfo{author}{A.~Hansen}, \bibinfo{author}{S.~Roux},
  \bibinfo{author}{A.~Aharony}, \bibinfo{author}{J.~Feder},
  \bibinfo{author}{T.~J{\o}ssang}, \bibinfo{author}{H.~H. Hardy},
\newblock \bibinfo{title}{{Real-Space Renormalization Estimates for Two-Phase
  Flow in Porous Media}},
\newblock \bibinfo{journal}{Transp. Porous Media} \bibinfo{volume}{29}
  (\bibinfo{year}{1997}) \bibinfo{pages}{247--279}.
\bibitem[{Wei et~al.(2019)Wei, Shen, Yang, Li, Di, and Ma}]{Wei2019}
\bibinfo{author}{S.~Wei}, \bibinfo{author}{J.~Shen}, \bibinfo{author}{W.~Yang},
  \bibinfo{author}{Z.~Li}, \bibinfo{author}{S.~Di}, \bibinfo{author}{C.~Ma},
\newblock \bibinfo{title}{{Application of the renormalization group approach
  for permeability estimation in digital rocks}},
\newblock \bibinfo{journal}{J. Pet. Sci. Eng.} \bibinfo{volume}{179}
  (\bibinfo{year}{2019}) \bibinfo{pages}{631--644}.
\bibitem[{Mavko et~al.(2003)Mavko, Mukerji, and Dvorkin}]{mavko2003rock}
\bibinfo{author}{G.~Mavko}, \bibinfo{author}{T.~Mukerji},
  \bibinfo{author}{J.~Dvorkin}, \bibinfo{title}{{The Rock Physics Handbook:
  Tools for Seismic Analysis of Porous Media}}, Stanford-Cambridge program,
  \bibinfo{publisher}{Cambridge University Press}, \bibinfo{year}{2003}.
\bibitem[{Ela(????)}]{Elastodict}
\bibinfo{title}{{http://www.geodict.de/Modules/Dicts/ElastoDict.php.}}
\bibitem[{Kabel and Andr{\"{a}}(2013)}]{Kabel2013}
\bibinfo{author}{M.~Kabel}, \bibinfo{author}{H.~Andr{\"{a}}},
\newblock \bibinfo{title}{{Fast numerical computation of effective elastic
  moduli of porous materials}},
\newblock \bibinfo{journal}{Rep. Fraunhofer ITWM} \bibinfo{volume}{224}
  (\bibinfo{year}{2013}) \bibinfo{pages}{1--16}.
\bibitem[{Kabel et~al.(2016)Kabel, Fliegener, and Schneider}]{Kabel2016}
\bibinfo{author}{M.~Kabel}, \bibinfo{author}{S.~Fliegener},
  \bibinfo{author}{M.~Schneider},
\newblock \bibinfo{title}{{Mixed boundary conditions for FFT-based
  homogenization at finite strains}},
\newblock \bibinfo{journal}{Comput. Mech.} \bibinfo{volume}{57}
  (\bibinfo{year}{2016}) \bibinfo{pages}{193--210}.
\bibitem[{de~Geus and Vondřejc(????)}]{Geus}
\bibinfo{author}{T.~W. de~Geus}, \bibinfo{author}{J.~Vondřejc},
  \bibinfo{title}{http://goosefft.geus.me/}
\bibitem[{de~Geus et~al.(2017)de~Geus, Vondřejc, Zeman, Peerlings, and
  Geers}]{DeGeus2017a}
\bibinfo{author}{T.~W. de~Geus}, \bibinfo{author}{J.~Vondřejc},
  \bibinfo{author}{J.~Zeman}, \bibinfo{author}{R.~H. Peerlings},
  \bibinfo{author}{M.~G. Geers},
\newblock \bibinfo{title}{{Finite strain FFT-based non-linear solvers made
  simple}},
\newblock \bibinfo{journal}{Comput. Methods Appl. Mech. Eng.}
  \bibinfo{volume}{318} (\bibinfo{year}{2017}) \bibinfo{pages}{412--430}.
\bibitem[{Zeman et~al.(2017)Zeman, de~Geus, Vondřejc, Peerlings, and
  Geers}]{Zeman2017}
\bibinfo{author}{J.~Zeman}, \bibinfo{author}{T.~W. de~Geus},
  \bibinfo{author}{J.~Vondřejc}, \bibinfo{author}{R.~H. Peerlings},
  \bibinfo{author}{M.~G. Geers},
\newblock \bibinfo{title}{{A finite element perspective on nonlinear FFT-based
  micromechanical simulations}},
\newblock \bibinfo{journal}{Int. J. Numer. Methods Eng.} \bibinfo{volume}{111}
  (\bibinfo{year}{2017}) \bibinfo{pages}{903--926}.
\bibitem[{Huet(1990)}]{Huet1990}
\bibinfo{author}{C.~Huet},
\newblock \bibinfo{title}{{Application of variational concepts to size effects
  in elastic heterogeneous bodies}},
\newblock \bibinfo{journal}{J. Mech. Phys. Solids} \bibinfo{volume}{38}
  (\bibinfo{year}{1990}) \bibinfo{pages}{813--841}.
\bibitem[{Hazanov and Huet(1994)}]{Hazanov1994}
\bibinfo{author}{S.~Hazanov}, \bibinfo{author}{C.~Huet},
\newblock \bibinfo{title}{{Order relationships for boundary conditions effect
  in heterogeneous bodies smaller than the representative volume}},
\newblock \bibinfo{journal}{J. Mech. Phys. Solids} \bibinfo{volume}{42}
  (\bibinfo{year}{1994}) \bibinfo{pages}{1995--2011}.
\bibitem[{Kanit et~al.(2003)Kanit, Forest, Galliet, Mounoury, and
  Jeulin}]{Kanit2003a}
\bibinfo{author}{T.~Kanit}, \bibinfo{author}{S.~Forest},
  \bibinfo{author}{I.~Galliet}, \bibinfo{author}{V.~Mounoury},
  \bibinfo{author}{D.~Jeulin},
\newblock \bibinfo{title}{{Determination of the size of the representative
  volume element for random composites: Statistical and numerical approach}},
\newblock \bibinfo{journal}{Int. J. Solids Struct.} \bibinfo{volume}{40}
  (\bibinfo{year}{2003}) \bibinfo{pages}{3647--3679}.
\bibitem[{Vondřejc et~al.(2015)Vondřejc, Zeman, and Marek}]{Vondrejc2015}
\bibinfo{author}{J.~Vondřejc}, \bibinfo{author}{J.~Zeman},
  \bibinfo{author}{I.~Marek},
\newblock \bibinfo{title}{{Guaranteed upper-lower bounds on homogenized
  properties by FFT-based Galerkin method}},
\newblock \bibinfo{journal}{Comput. Methods Appl. Mech. Eng.}
  \bibinfo{volume}{297} (\bibinfo{year}{2015}) \bibinfo{pages}{258--291}.
\bibitem[{Moulinec and Suquet(1994)}]{Moulinec1994}
\bibinfo{author}{H.~Moulinec}, \bibinfo{author}{P.~Suquet},
\newblock \bibinfo{title}{{A fast numerical method for computing the linear and
  nonlinear mechanical properties of composites}},
\newblock \bibinfo{journal}{Mech. solids}  (\bibinfo{year}{1994}).
\bibitem[{Moulinec and Suquet(1998)}]{Moulinec1998a}
\bibinfo{author}{H.~Moulinec}, \bibinfo{author}{P.~Suquet},
\newblock \bibinfo{title}{{A numerical method for computing the overall
  response of nonlinear composites with complex microstructure}},
\newblock \bibinfo{journal}{Comput. Methods Appl. Mech. Eng.}
  \bibinfo{volume}{157} (\bibinfo{year}{1998}) \bibinfo{pages}{69--94}.
\bibitem[{Kabel et~al.(2014)Kabel, B{\"{o}}hlke, and Schneider}]{Kabel2014}
\bibinfo{author}{M.~Kabel}, \bibinfo{author}{T.~B{\"{o}}hlke},
  \bibinfo{author}{M.~Schneider},
\newblock \bibinfo{title}{{Efficient fixed point and Newton–Krylov solvers
  for FFT-based homogenization of elasticity at large deformations}},
\newblock \bibinfo{journal}{Comput. Mech.} \bibinfo{volume}{54}
  (\bibinfo{year}{2014}) \bibinfo{pages}{1497--1514}.
\bibitem[{Vondřejc et~al.(2014)Vondřejc, Zeman, and Marek}]{Vondrejc2014a}
\bibinfo{author}{J.~Vondřejc}, \bibinfo{author}{J.~Zeman},
  \bibinfo{author}{I.~Marek},
\newblock \bibinfo{title}{{An FFT-based Galerkin method for homogenization of
  periodic media}},
\newblock \bibinfo{journal}{Comput. Math. with Appl.} \bibinfo{volume}{68}
  (\bibinfo{year}{2014}) \bibinfo{pages}{156--173}.
\bibitem[{Leute et~al.(2022)Leute, Ladeck{\'{y}}, Falsafi, J{\"{o}}dicke,
  Pultarov{\'{a}}, Zeman, Junge, and Pastewka}]{Leute2022}
\bibinfo{author}{R.~J. Leute}, \bibinfo{author}{M.~Ladeck{\'{y}}},
  \bibinfo{author}{A.~Falsafi}, \bibinfo{author}{I.~J{\"{o}}dicke},
  \bibinfo{author}{I.~Pultarov{\'{a}}}, \bibinfo{author}{J.~Zeman},
  \bibinfo{author}{T.~Junge}, \bibinfo{author}{L.~Pastewka},
\newblock \bibinfo{title}{{Elimination of ringing artifacts by finite-element
  projection in FFT-based homogenization}},
\newblock \bibinfo{journal}{J. Comput. Phys.} \bibinfo{volume}{453}
  (\bibinfo{year}{2022}) \bibinfo{pages}{110931}.
\bibitem[{Schneider(2021)}]{Schneider2021}
\bibinfo{author}{M.~Schneider},
\newblock \bibinfo{title}{{A review of nonlinear FFT-based computational
  homogenization methods}},
\newblock \bibinfo{journal}{Acta Mech.} \bibinfo{volume}{232}
  (\bibinfo{year}{2021}) \bibinfo{pages}{2051--2100}.
\bibitem[{Lucarini et~al.(2022)Lucarini, Upadhyay, and Segurado}]{Lucarini2022}
\bibinfo{author}{S.~Lucarini}, \bibinfo{author}{M.~V. Upadhyay},
  \bibinfo{author}{J.~Segurado},
\newblock \bibinfo{title}{{FFT based approaches in micromechanics:
  fundamentals, methods and applications}},
\newblock \bibinfo{journal}{Model. Simul. Mater. Sci. Eng.}
  \bibinfo{volume}{30} (\bibinfo{year}{2022}) \bibinfo{pages}{023002}.

\end{thebibliography}

\end{document}